\documentclass{aa}
\usepackage{txfonts}
\usepackage{graphics}
\usepackage{latexsym}
\begin{document}

\title{
An atlas of line profile studies\\ for SU UMa type cataclysmic 
variables\thanks{Partly based on observations at ESO telescopes}\fnmsep
\thanks{Figs.\ \ref{akdd_fig}, \ref{akdop_fig}, and \ref{aqdd_fig}
to \ref{hsdop_fig} are only available in electronic form at 
http://www.edpsciences.org}
}

\author{
C. Tappert\inst{1}
\and
R. E. Mennickent\inst{1}
\and
J. Arenas\inst{1}
\and
K. Matsumoto\inst{2}
\and
R. W. Hanuschik\inst{3}
}

\institute{
Universidad de Concepci\'on, Departamento de F\'{\i}sica, Casilla 160-C, 
Concepci\'on, Chile
\and
Graduate School of Natural Science and Technology, Okayama University, 
Okayama 700-8530, Japan
\and
ESO, Karl-Schwarzschild-Str. 2, D-85748 Garching, Germany
}

\offprints{C. Tappert \email{claus@gemini.cfm.udec.cl}}

\date{Received / Accepted}

\titlerunning{An atlas of line profile studies for SU UMa stars}

\authorrunning{Tappert et al.}

\abstract{
We present H$\alpha$ line-profile analyses for the seven SU UMa type dwarf 
novae AK Cnc, WX Cet, AQ Eri, VW Hyi, RZ Leo, TU Men, and HS Vir. All data sets
are treated in the same manner, applying a sequence of techniques for each 
system. The basic ingredients of this sequence are the diagnostic diagram to 
determine the zero point of the orbital phase, and Doppler tomography to 
visualise the emission distribution. We furthermore introduce a new 
qualitative way of to evaluate the Doppler fit, by comparing the line profile 
of the reconstructed with the original spectrum in the form of the $V/R$ plot. 
We present the results of the analysis in the compact form of an atlas, 
allowing a direct comparison of the emission distribution in our targets. 
Although most of the data sets were not taken with the intention of a 
line-profile analysis, we obtain significant results and are able to indicate 
the type of the additional emission in these systems. Our objects should have 
in principle very similar physical properties, i.e.\ they cover only a small 
range in orbital periods, mass ratios, and mass-transfer rates. Nevertheless, 
we find a large variety of phenomena both with respect to the individual 
systems and also within individual data sets of the same object. This includes 
`canonical' additional emission components from the secondary star and the 
bright spot, but also emission from the leading side of the accretion disc.
}

\maketitle

%
%________________________________________________________________

\section{Introduction}

Cataclysmic variables (CVs) are close interacting binaries with a white
dwarf as primary and a late-type main-sequence star as secondary component. 
The latter fills its critical Roche lobe and thus enables mass transfer through
the inner Lagrangian point into the Roche volume of the primary. In the absence
of strong magnetic fields, the gas stream from the secondary star dissipates 
and forms an accretion disc around the white dwarf over which the mass transfer 
takes place.

The accretion disc is the dominant light source in these systems. It is 
also the primary origin of the strong emission lines especially of the 
H and He series, which are characteristic for a CV spectrum (e.g.\ Smak 
\cite{smak81b}; Horne \& Marsh \cite{hornmars86}; Horne \cite{horn95}). In most
CVs, the disc is too bright to allow the detection of the absorption features 
of the primary and the secondary. The emission lines thus represent the major 
source of information on the orbital motion. Furthermore, the assumption of a 
Keplerian velocity law allows the derivation of several system parameters like 
masses and inclination. These parameters are, of course, vital for an 
understanding of the physics in these systems, and furthermore represent the 
only means to test theoretical models on CV evolution (see, e.g.\ Warner 
\cite{warn95} for a comprehensive overview on CVs).

This approach is only feasible if the emission lines originate in a
region moving with the same mean velocity vector as the primary star.
Only in this case do they track
its orbital motion. However, several examples show that, additionally to the
disc-borne emission, other emission components distort the originally symmetric
line profile (e.g.\ Stover \cite{stov81a}). The most prominent sources of
this emission are the region around the bright spot (the place where the gas 
stream impacts on the disc), the gas stream from the secondary star, or an 
irradiated part on the surface of the latter, but also apparently enhanced 
emission regions in the accretion disc, which, in a number of cases, make a 
physical explanation difficult (Tappert \cite{tapp99}; North et al.\ 
\cite{nort+01}). All these regions have in common that they are `isolated' in 
the sense that they are not centred symmetrically on the white dwarf. 
Consequently, they have both different velocity amplitudes and phasing with 
respect to the primary. The radial velocity parameters of the combined line 
profile as a whole thus will not reflect the motion of the white dwarf, leading
to erroneous system parameters. We will call these asymmetric emission regions 
in the following `isolated emission sources' (IES).

Fortunately, in most cases, the distorting components are confined to the line 
centre.\footnote{The high velocity wings in SW Sex systems are an important 
exception (e.g.\ Hoard et al.\ \cite{hoar+00}, and references therein).} A 
careful examination of the line profile is thus usually able to separate the 
influences of additional components and allows the determination of the system 
parameters with some confidence. Furthermore, a recent statistical study showed
that the additional components represent more than just a mere obstacle in
the derivation of the system parameters, but in principle provide valuable
information on the physical processes in CVs (Tappert \& Hanuschik 
\cite{tapphanu01}). Last, not least, the discovery of spiral shocks in CV discs
by the means of Doppler tomography (Steeghs et al.\ \cite{stee+97}) proves the 
importance of line profile analyses for the understanding of accretion physics.

Kaitchuck et al.\ (\cite{kait+94}) were the first to re-analyse a large sample 
of `old', i.e.\ already elsewhere published, CV spectra with a newly developed
technique. Their atlas of Doppler maps allowed a direct comparison of the 
various types of emission distribution. The atlas included 18 systems, 3
of them with an orbital period shorter than 3 hours. 

In a similar spirit, the scope of the present paper is to apply a sequence of
techniques of line profile analysis to a sample of CVs. Our concentration on 
short-period dwarf novae allows us to study the variety of phenomena presented 
by systems which can be assumed to have very similar physical properties, 
i.e.\ a low mass ratio $q = M_2/M_1 \le 0.3$ and low 
mass-transfer rates of the
order $\dot{M} \sim 10^{-11} M_{\odot}/\mathrm{yr}$. Furthermore, according to 
standard CV evolution theory, low mass transfer rates imply optically thin 
discs. These are strong line emitters and could reveal structural details more 
easily than optically thick discs. 

The present atlas also provides a significant increase of Doppler maps for
short-period CVs, whose number of available line-profile analyses is in
strong mismatch with respect to the number of known systems, compared with
long-period CVs (Tappert \& Hanuschik \cite{tapphanu01}). This is basically
due to the fact that for these faint systems the orbital period is easily
accessible for small telescopes through the superhump phenomenon, while 
detailed studies need large telescopes. The latter therefore usually 
concentrate on a few `interesting' objects, like \object{WZ Sge} (e.g.,
Skidmore et al.\ \cite{skid+00}; and references therein). Our present sample
instead is not part of this `selection effect', as none of the systems has 
(yet) been the subject of a more detailed investigation involving the 
necessary large telescopes. It therefore adds to a more `unbiased' view of
the properties of SU UMa stars.

Following our goal of direct comparison, we restrict our analysis to the 
H$\alpha$ profile, which also represents the most accessible emission line.

\section{The sample}

\begin{table*}
\caption[]{General properties of the data. The dates of observations and the 
corresponding data sets (numbers usually refer to individual nights; for
exceptions see text) are presented in Col.\ 2 and 3, respectively. Column
4 gives the total number of spectra available for the analysis, Col.\ 5 their 
spectral resolution (FWHM), Col.\ 6 the telescope used for the observations, 
and the last column lists references concerning the previous publication of the
data sets.}
\label{data_tab}
\begin{center} 
\begin{tabular}{l l l l l l l} 
\hline
star & dates & sets & $n_{\rm sp}$ & $\Delta\lambda$ [{\AA}] 
& telescope & reference\\ 
\hline 
\object{AK Cnc} & 1995-03-19, 21, 22        & 1--3           & 40 & 4.5 
& 2.5m LCO & Arenas \& Mennickent (\cite{arenmenn98})\\
\object{WX Cet} & 1992-12-07, 08, 10        & 1--3           & 37 & 2.0 
& 2.2m ESO/MPI & Mennickent (\cite{menn94})\\
\object{AQ Eri} & 1992-12-11                & 1(a--c)        & 44 & 2.0 
& 2.2m ESO/MPI & Mennickent (\cite{menn95b})\\
\object{VW Hyi} & 1994-10-15 to 17          & 1--3           & 64 & 1.1
& 1.9m SAAO & Tappert (\cite{tapp99}) \\
\object{RZ Leo} & 1995-02-10, 11, 03-21, 23 & 1a, 1b, 2a, 2b & 46 & 6.0, 4.5 
& 2.2m ESO/MPI, 2.5m LCO & Mennickent \& Tappert (\cite{menntapp01})\\
\object{TU Men} & 1992-12-09, 10, 12, 14    & 1--4           & 67 & 2.0 
& 2.2m ESO/MPI & Mennickent (\cite{menn95a})\\
\object{HS Vir} & 1998-05-30                & 1              & 34 & 2.5 
& 3.5m NTT & Mennickent et al.\ (\cite{menn+99b})\\
\hline
\end{tabular} 
\end{center} 
\end{table*} 

The systems in question are the seven SU UMa type dwarf novae 
AK Cnc, WX Cet, AQ Eri, VW Hyi, RZ Leo, TU Men, and HS Vir. 
With the exception of VW Hyi, the data were not taken with the intention to 
perform a line-profile analysis, but to determine the orbital period. 
The observational background and the properties of the data have been published
elsewhere, together with the results of the initial analysis. In Table 
\ref{data_tab} we therefore present only a summary of the basic parameters of 
the data sets, and refer the reader to the respective publications listed in 
the table for further details.

The available data on HS Vir actually consists of more than the one set covered
here. However, the system showed large, apparently erratic, variations from
one set to another, and also within individual data sets. A more detailed
analysis is required in this case, which is beyond the scope of the present 
paper, since it aims at a uniform treatment of the data. We have therefore 
included only the `most well behaved' data set of this object as an example.

\section{Analysis sequence\label{anal_sec}}

The following sequence of analysis techniques is applied to each data set:
\medskip\\

\noindent
{\sl 1. Visual inspection:}
Apart from the changes due to the orbital motion and/or to additional
emission sources, the emission-line profile is also affected by luminosity
changes of the system. Roughly, the strength of the disc-borne emission 
lines is expected to decrease with increasing disc brightness, due to the 
increasing optical thickness of the accretion disc. Additional emission 
components, however, might be affected quite differently by such a variation, 
so that their influence becomes stronger (e.g., emission from the secondary 
star) or weaker (e.g., emission from the bright spot). Line profiles of the 
same CV in different states thus usually show quite different phenomena 
(compare, e.g., Mart\'{\i}nez-Pais et al.\ \cite{mart+94}, \cite{mart+96}, for 
\object{SS Cyg} in quiescence and outburst, respectively). A visual comparison 
of line profiles of different data sets therefore provides a first hint 
as to whether a combination of these data is possible in order to increase the 
phase resolution, or if each set has to be treated individually.

Similarly, in high inclination systems, the brightness variation due to the 
orbital motion can affect the line profile at specific orbital phases. This is 
especially true for eclipsing systems, where also the line emission source is 
partly or completely obscured (e.g.\ Kaitchuck et al.\ \cite{kait+98}). These 
parts of the data sets usually have to be excluded from the line profile 
analysis.
\medskip\\

\noindent
{\sl 2. Diagnostic diagram:}
This method makes use of the 
double-Gaussian convolution with the line profile (Schneider \& Young 
\cite{schnyoun80}; Shafter \cite{shaf83b}), which allows to measure radial
velocities of different parts of the line. These velocities $v_r(\varphi)$ can 
be fitted with a sinusoidal function
\begin{equation}
v_r(\varphi) = \gamma - K_\mathrm{em} \sin (2 \pi (\varphi-\varphi_0)), 
\label{radvec_eq}
\end{equation}
to yield the systemic velocity $\gamma$, the semi-amplitude $K_\mathrm{em}$, 
and the fiducial phase corresponding to superior conjunction of the primary 
star $\varphi_0$. When these parameters are plotted as a function
of the separation $d$ of the two Gaussians, a so-called diagnostic diagram is 
obtained, which illustrates the variation of the parameters as different parts 
of the line are measured (Shafter \cite{shaf83a}). In order to evaluate the 
significance of the resulting variation, the corresponding error of the 
semi-amplitude $\sigma(K_\mathrm{em})/K_\mathrm{em}$ is included in the plot. A
sharp increase of this value at large separations indicates the point beyond 
which the noise begins to dominate the signal from the line wings and the 
parameters become unreliable. The values belonging to the largest separation 
$d_{\rm max}$ before this happens are usually chosen as the best possible 
approximation to the correct ones, so that $K_\mathrm{em} = K_1$, i.e.\ the 
true semi-amplitude of the orbital motion of the primary.

A second criterion is based on the expectation that the parameters should
(again) approach constant values at large separations when additional emission
sources cease to influence the line profile. This is usually the stronger
criterion, as the first one is often ambiguous.

The diagnostic diagram furthermore allows us to fix the fiducial phase, 
which is a prerequisite for the evaluation of the subsequent
Doppler mapping.
\medskip\\

\noindent
{\sl 3. Doppler tomography:}
%This technique, introduced by Marsh \& Horne (\cite{marshorn88}) to the
%realm of interacting binaries, back-projects the one-dimensional snapshots
%represented by a spectrum taken at a specific orbital phase onto a 
%two-dimensional velocity grid, thus obtaining an image of emission from the
%accretion disc in velocity space (see Marsh \cite{mars01} for a recent review). 

This technique, introduced by Marsh \& Horne (\cite{marshorn88}) to the
realm of interacting binaries, interprets the one-dimensional snapshots
represented by a spectrum taken at a specific orbital phase as projections
of a two-dimensional emission distribution. The latter is reconstructed as 
a map in velocity space via an inversion of these projections
(see Marsh \cite{mars01} for a recent review).
 
In this work, the implementation of Spruit (\cite{spru98}) has been used.
We have replaced the original IDL routines by a corresponding MIDAS interface, 
but still use the FORTRAN core program (version 2.3.1), to run the computation 
on a Linux PC. The input spectra have first been continuum subtracted, the
extracted line profiles have then been normalised in flux (in order to avoid
too small numbers for the Doppler routine, the total line flux has been set to
100) to minimise artefacts due to intensity variations. A smearing kernel of 2 
pixels was used to compute the Doppler maps, which were subsequently smoothed 
with a $3\times3$ pixel ($11\times11$ in the case of VW Hyi) averaging filter 
for noise reduction. Tests with larger kernels showed that the visual 
information content of the resulting map basically remains the same, but that 
the profile comparison (see step 4) suffered from too strong smoothing. A 
kernel size of 1 pixel, on the other hand, in almost all cases led to infinite 
iteration loops due to the noise becoming dominant. All other input parameters 
for the Doppler routine were set to default values (Spruit \cite{spru98}).

The $\gamma$ velocity resulting from the diagnostic diagram was taken as a
first guess for correcting the input line to its rest wavelength. It was then
iteratively adjusted by comparing the reconstructed and original average 
spectra.

The resulting Doppler maps are orientated as usual, i.e.\ 
the fiducial phase is at $v_x$ = 0 and $v_y >$ 0
and increases in clockwise direction. If the phase corresponds to the orbital 
motion this results in the white dwarf being located at ($v_x$,$v_y$) = 
(0,$-K_1$) and the secondary star at (0,$K_2$).
\medskip\\

\noindent
{\sl 4. $V/R$-plot:}
Perhaps the simplest way to examine a line profile for asymmetry is
to compare its blue part to its red one. This has been done in the past for
double-peaked profiles by computing the ratio between the intensity of the
blue (violet) peak and the one of the red peak, the so-called $V/R$ ratio (e.g.,
Mennickent \cite{menn94}). In order to apply this method also to single-peaked
profiles, one can define a different, but equivalent, $V/R$ as (Tappert 
\cite{tapp99})
\begin{equation}
 V/R = \log \frac{F(V)}{F(R)}~,
 \label{vrdiff_eq}
\end{equation}
where $F(V)$ is not the peak intensity but the flux of the blue part and
$F(R)$ the corresponding one of the red part. The point which separates both 
halves, $\lambda_c$, is chosen as the centre of the line flanks at a specific
intensity value which has to be low enough to avoid being affected by the 
central valley in double-peaked profiles, and high enough to avoid the 
noise-dominated extreme line wings. The sigma of the individual $V/R$ value
is computed by a Monte Carlo simulation which adds a random value, uniformly 
distributed within an interval determined by the S/N, to each data point in the
spectrum, and measures the $V/R$ for a thousand of such newly computed data 
sets.

We here use the $V/R$ plot to compare the original spectrum with the 
reconstructed one from the Doppler map. In this way it can be quantitatively
evaluated if the Doppler tomography has been able to reproduce the general
shape of the individual line profiles.\\ 
\medskip

\noindent
In the atlas presented in Appendix \ref{atlas_sec}, we furthermore include 
two-dimensional greyscale plots of the original and the difference data, the 
latter being computed by subtracting the reconstructed spectrum from the 
original one. In this way, also possible intensity deficits (black features in 
the plot) or excesses (white features) of the reconstructed data become 
visually clear.

\section{Results\label{res_sec}}

\subsection{Splitting of data sets}

For all objects, the analysis methods were first conducted on each individual
data set (i.e., corresponding to one night of observation; see Table 
\ref{data_tab}). Where no significant differences were found, the analysis was 
repeated on the combined set in order to minimise noise.

The data of AQ Eri represent a special case. Here, the line profiles proved
to vary both in strength and in shape on non-orbital timescales (compare
especially phases 0.1 and 0.8 in the upper plot of Fig.\ \ref{aqdd_fig}).
We therefore split the set in three parts 1a, 1b, 1c, with each set covering
one full orbit. Set 1b overlaps with the other two (it includes 4 spectra
from set 1a and 3 from set 1c), but sets 1a and 1c are independent of
each other.

The appearance of the line profiles of HS Vir (upper plot of Fig.\ 
\ref{hsdd_fig}) suggests a similar treatment. However, here the phase
distribution of the data did not allow a division into independent sets
which would cover a complete orbit. We conducted the analyses for different
subsets, but did not find any significant differences with respect to results 
or residuals.

\subsection{Diagnostic diagrams}

\begin{table}
\caption[]{Radial-velocity parameters for H$\alpha$ as derived by the analysis 
of the diagnostic diagram. Col.\ 2 gives the FWHM of the Gaussians used for the
diagram, Col.\ 3 the critical separation, Col.\ 4 and 5 contain the
resulting semi-amplitude and the error of the zero phase, respectively.
The $\gamma$ parameter is not listed, as it reflects more the wavelength
calibration process than a real physical parameter.} 
\label{ddpar_tab}
\begin{tabular}{lclrll}
\hline
data set & FWHM & $d_{\rm max}$ & \multicolumn{1}{c}{$K_\mathrm{em}$} & 
 $\sigma(\varphi_0)$ & Fig.\\
 & [{\AA}] & [{\AA}] & [km/s] & [orbits] & \\
\hline
AK Cnc   & 4 & 34 & 42(09) & 0.036 & \ref{akdd_fig} \\
WX Cet   & 4 & 52 & 59(18) & 0.061 & \ref{wxdd_fig} \\
AQ Eri   & 2 & 44 & 36(09) & 0.042 & \ref{aqdd_fig} \\
VW Hyi   & 4 & 40 & 38(09) & 0.039 & \ref{vwdd_fig} \\
RZ Leo 1 & 4 & 66 & 49(13) & 0.042 & -- \\
RZ Leo 2 & 4 & 62 & 51(15) & 0.050 & \ref{rzdd_fig} \\
TU Men   & 2 & 48 & 84(09) & 0.017 & \ref{tudd_fig}\\
HS Vir   & 2 & 64 & 98(09) & 0.017 & \ref{hsdd_fig} \\
\hline
\end{tabular}
\end{table}

The diagnostic diagrams proved to be robust with respect to the differences
between the individual data sets. With the exception of RZ Leo, where the
insufficient precision of the orbital period does not allow for an 
interpolation of the phases, we therefore used the combined data sets.

In the present application of the diagnostic diagram we are mostly 
interested in determining the zero point of the orbital phase, on which
depends the interpretation of the subsequent Doppler maps. For the
majority of the objects in our sample, this choice was unambiguous, as after
clearing distortions due to structure in the line core this parameter remained 
constant over a large range of Gaussian separations. We therefore based our 
choice of the maximum separation $d_{\rm max}$ on the stability of the other 
parameters (of $\gamma$ in the case of AK Cnc and WX Cet, $\gamma$ and 
$K_\mathrm{em}$ for TU Men) where the noise parameter 
$\sigma(K_\mathrm{em})/K_\mathrm{em}$ appeared not to be reliable, on a 
combination of the noise and the constancy criteria (AQ Eri), or (for HS Vir) 
only on the noise behaviour. In the cases of VW Hyi and RZ Leo, $\varphi_0$
does not reach constancy within the allowed range with respect to the 
noise criterion. We here chose the values corresponding to the largest 
separation within this range. Judging from the appearance of the variation 
of $\varphi_0$ at the separations before this point (slight monotonic decline),
we do not expect the error in the determination of the fiducial phase 
to be larger than $\sim$0.1 orbits.

For completeness, we have listed the parameters corresponding to our choice
of $d_{\rm max}$ in Table \ref{ddpar_tab}.

\subsection{Doppler maps}

The general appearance of the Doppler maps is a rather noisy one. This is
due to the fact that the S/N of the data could not be improved in the majority 
of our systems by co-adding phases of different orbits because of the observed 
short- and long-term variations. For the same reason also the phase coverage
often stays incomplete and/or non-uniform. All this gives rise to artefacts in 
the Doppler maps, and one has therefore to be careful not to over-interpret 
certain features.

Nevertheless, the tests by the means of the 
trailed spectrograms and the $V/R$ plots show that the
reconstructed spectra produced by the Doppler fitting agree fairly well with 
the original data, with the exception of HS Vir (see Sect.\ \ref{hs_sec}).
Differences in the $V/R$ values (e.g., lower plots for WX Cet, Figs.\ 
\ref{wx1dop_fig} and \ref{wx3dop_fig}; RZ Leo, Fig.\ \ref{rz2dop_fig}) can 
largely be attributed to noise in the original spectra and to the fact that due
to the Doppler smoothing (or smearing) the S/N in the reconstructed spectra is 
often up to 5 times higher. We therefore conclude that the line profiles 
themselves have been well reproduced. As shown by the 
trailed spectrograms, in 
general the same can be said for the intensity behaviour, with a few notable 
exceptions which are discussed below for the respective systems.

\subsubsection{AK Cnc}

The Doppler map shows a symmetric emission distribution (Fig.\ 
\ref{akdop_fig}). AK Cnc therefore appears to be one of the few systems which 
show no presence of any IES; the survey of Tappert \& Hanuschik 
(\cite{tapphanu01}) found only 4 such objects in a sample of 68 CVs (not 
including AK Cnc). Unfortunately, the system was not observed in quiescence,
but in decline from outburst. It thus remains unclear if the absence of IES is 
caused by a still optically thick disc (i.e.\ if this is only a temporary 
absence), or simply due to the low inclination of the system.

\subsubsection{WX Cet}

\begin{figure}
\rotatebox{270}{\resizebox{6cm}{!}{\includegraphics{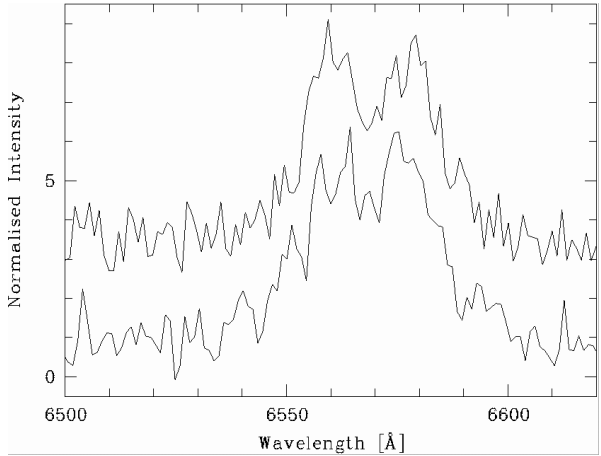}}}
\caption[]{Example of an original continuum-normalised spectrum of the WX Cet 
data set (bottom) and its corresponding model spectrum (top). The latter has
been displaced vertically by 2.5 intensity units.}
\label{wxmodprof_fig}
\end{figure}

\begin{figure}
\resizebox{7cm}{!}{\includegraphics{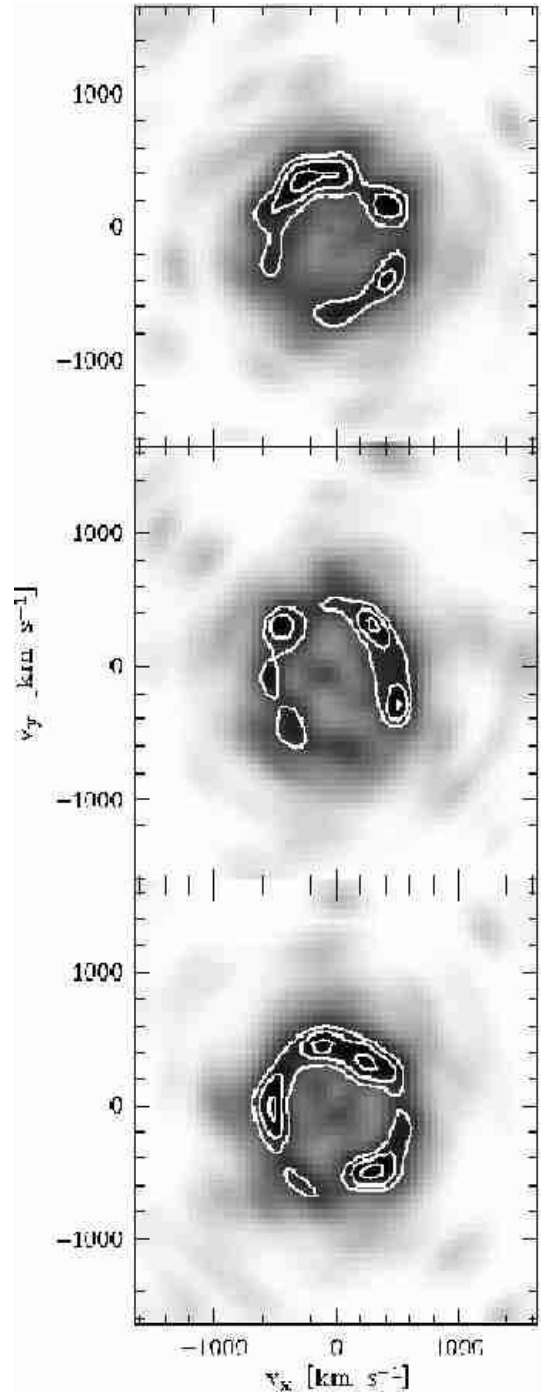}}
\caption[]{Doppler maps for the model spectra with respect to the WX Cet 
data sets 1--3 (top to bottom).}
\label{wxmoddop_fig}
\end{figure}

The Doppler maps for the WX Cet are amongst those with the most noisy 
appearance (Figs.\ \ref{wx1dop_fig}-\ref{wx3dop_fig}). The low S/N of the data 
is probably also responsible for the fact that the clear signature of 
disc emission, i.e.\ a ring-shaped emission distribution centred on the 
primary, is not seen, although the line profiles are double-peaked
(Fig.\ \ref{wxdd_fig}). Instead all maps show a rather spotty emission 
distribution, however, with a maximum on the leading side 
(i.e.\ $v_x > 0$) as a common feature.

The low S/N and the non-uniform phase distribution make the WX Cet data sets
the ideal test case to examine the significance of the features in the 
Doppler map. For this, we established a set of model spectra using the line 
profile prescription by Smak (\cite{smak81b}). We adjusted the model
parameters until a visually good agreement with the shape and quality of the 
original WX Cet data was reached. The model data were displaced in wavelength
according to Eq.\ (\ref{radvec_eq}), using the phases of the WX Cet data and 
the parameters obtained by the application of the diagnostic diagram ($K_1$ = 
59 km s$^{-1}$) and the Doppler tomography ($\gamma$ = 260 km s$^{-1}$).
Fig.\ \ref{wxmodprof_fig} gives an example of the resulting model spectrum. 

These spectra represent a symmetric disc emission, without any additional 
component. The corresponding Doppler maps (Fig.\ \ref{wxmoddop_fig}), however,
show the same spotty appearance as for the original data set, with the
ring-shaped disc signature being only slightly more evident. The low 
intensity arcs and blobs in many of the Doppler maps (see also the VW Hyi data,
especially Fig.\ \ref{vw2dop_fig}) can also be explained by the data
quality. It even appears not improbable that the individual maps of WX Cet 
could be simply reproduced by noisy spectra without the need to invoke an 
additional component (compare especially the map for model set 2 with the WX 
Cet data). However, the fact that all WX Cet maps show an emission maximum on 
the leading side of the disc indicates that this represents a real feature.

\subsubsection{AQ Eri\label{aqdop_sec}}

The maps for the individual data sets show significant differences 
(Figs.\ \ref{aq1adop_fig}-\ref{aq1cdop_fig}), with an emission maximum close to
the possible position of the secondary star 
at $(0, K_2)$ moving to a position in the lower 
left quadrant of the Doppler map. However, the important difference to the 
behaviour in other systems (e.g.\ VW Hyi) is that this evolution
takes place over $\sim$2.5 orbits, i.e.\ on very short timescales. It is 
therefore doubtful if the maps in this case are representative of the
emission distribution at all, or if the timescales of the profile variations
in AQ Eri are shorter than even one orbit. The fact that the comparison 
between the original and the reconstructed data shows evidence of systematic
differences (especially the grey scale plot for set 1b and the $V/R$ plot for
set 1c), seems to indicate the latter.

\subsubsection{VW Hyi\label{vwdop_sec}}

\begin{figure}
\rotatebox{270}{\resizebox{5.7cm}{!}{\includegraphics{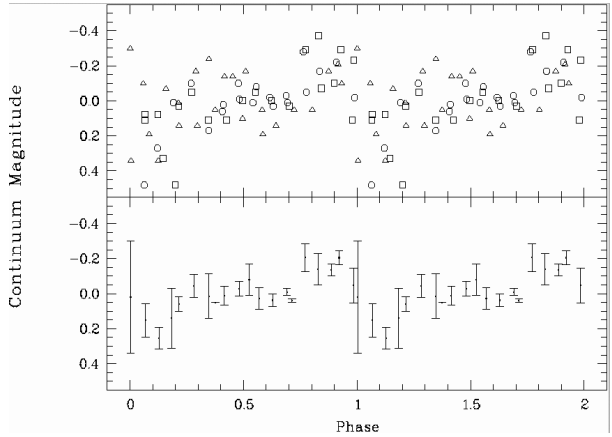}}}
\caption[]{The H$\alpha$ continuum magnitude of the VW Hyi data as a function 
of phase. The upper plot has been composed out of data sets 1 ($\circ$), 2 
($\bigtriangleup$), and 3 ($\Box$), with the respective average values 
subtracted. The lower plot gives those three data sets averaged into bins of 
0.05 orbital phases, with the error bars referring to $\pm 1 \sigma$ deviation
of the averaging process. The data are repeated to show two orbits.}
\label{vwcmag_fig}
\end{figure}

\begin{figure}
\rotatebox{270}{\resizebox{5.9cm}{!}{\includegraphics{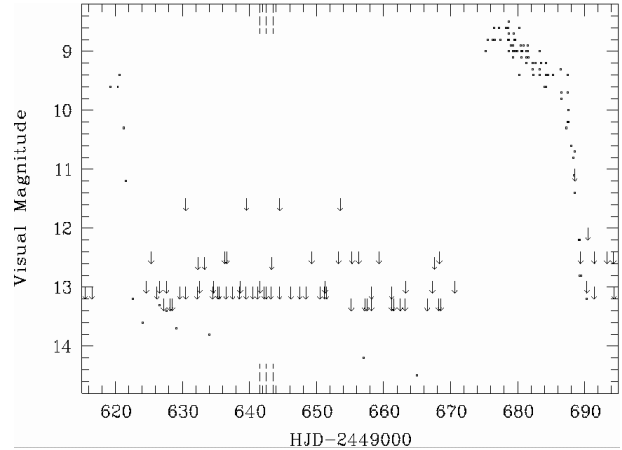}}}
\caption[]{Long-term light curve for VW Hyi based on AAVSO observations
(Mattei \cite{matt99}, private communication). Upper brightness limits are
marked by arrows. The dates of the spectroscopic measurements are indicated by
the dashed lines.}
\label{vwlt_fig}
\end{figure}

The maps show an evolution of the emission distribution from a combined
secondary star / gas stream location in set 1 to a dominating bright spot
in set 3 (Figs.\ \ref{vw1dop_fig}-\ref{vw3dop_fig}). The ring-shaped disc
signature is present, but -- like in WX Cet -- distorted due to the low S/N.

While the $V/R$ plot comparison of the original and the reconstructed data are 
in good agreement with each other, there are some systematic differences to be 
seen in the 
trailed spectrograms especially for set 1. This concerns a narrow 
emission component that is stronger in the reconstructed data from phases 0.7 
to 1.2 and weaker during the other phases. We attribute this to an occultation
effect of this emission component, the latter coming most probably from the
gas stream or the secondary star (which shows us its non-irradiated backside 
around phase 0.0). This violates one of the prerequisites for Doppler mapping, 
specifically that all emission sources have to be visible at all phases. We 
find a similar phenomenon for TU Men (Section \ref{tudop_sec}).

While we did not detect any systematic variation in the H$\alpha$ equivalent
widths, the continuum -- i.e.\ the optically thick emission -- shows a
hump at phases 0.75-1.0 -- probably from the gas stream --, followed by a 
shallow eclipse up to phase 1.2 (Fig.\ \ref{vwcmag_fig}).\footnote{The 
uncertainties in the determination of the continuum (which was measured with
a graphics cursor) are hard to quantify, but the fact that all three data sets
show the same basic features at identical phases proves the reality of this
variation.} Although the behaviour of the continuum 
may be different from 
that of the emission line, this indicates that the inclination of
the system is high enough for occultation effects. Also the fact that the 
differences in the 
trailed spectrograms decrease from set 1 to set 3, while at the 
same time the Doppler maps show that emission from the secondary star 
becomes less dominant in this sequence, supports
the interpretation that the narrow emission component originates from the
secondary star and is occulted during the corresponding phases.

The data presented here have been taken 21 days after maximum of a normal
outburst, with the system apparently being in a long, slow decline 
(Fig.\ \ref{vwlt_fig}). The phenomena reported here might therefore either
represent a continuous evolution from the outburst to the quiescent state,
or a short-term fluctuation (timescale of a few days).

\subsubsection{RZ Leo}

Both data sets yield very similar emission distributions in the form of two
maxima on the leading and on the trailing side of the disc (Figs.\ 
\ref{rz1dop_fig} and \ref{rz2dop_fig}), with the latter maximum being the 
stronger one. The intensity difference appears to be more pronounced in 
data set 2, but it remains unclear if this variation is significant or due
to noise. 

In view of the spotty appearance of other Doppler maps in this atlas (WX Cet, 
VW Hyi) it is justified to ask if the spots in RZ Leo might be artefacts as
well. However, in this case, also the trailed spectrograms show the
unambiguous presence of an additional component. At least the spectrogram for 
data set 1 furthermore shows evidence in the form of a double wave for a 
second component. While the second spot (on the leading side) in data set 2
might therefore be questionable, in set 1 the strength of this feature appears 
to exclude an artificial origin.

\subsubsection{TU Men\label{tudop_sec}}

The Doppler map of TU Men shows a strong emission feature from the secondary
star and otherwise pure disc emission (Fig.\ \ref{tudop_fig}). The differences
between the reconstructed and the original data which can be seen in the grey 
scale plots affect the same phases as in the case of VW Hyi (Section
\ref{vwdop_sec}). 
Also the explanation is the same: if an emission feature is not visible
at all phases (i.e.\ in violation of one of the prerequisites for Doppler
tomography), the Doppler fit will smear the signal over all phases. The feature
will therefore be artificially enhanced during the phases when it is absent
in the original data (here: phases 0.8--1.2), and it will be reconstructed
with diminished strength in the phases where it is visible 
(here: 0.3--0.7).\footnote{An extension of the Doppler tomography method to
account for emission components with variable strength has been very recently
proposed by Steeghs (\cite{stee03}).}

However, while for TU Men the residuals are more positive than negative, 
the VW Hyi data set 1 shows the opposite behaviour. In other words, in TU Men 
the IES feature is diminished rather than enhanced, while in VW Hyi the 
enhancement is stronger.

A possible reason could be a much stronger obscuration in VW Hyi. Indeed the 
differences in the strength of the emission feature at the respective phases 
appear to be more pronounced in VW Hyi than in TU Men. Unfortunately, the
TU Men data had not been flux-calibrated so that the strength of a possible
obscuration effect in TU Men cannot be compared to that in VW Hyi.

\subsubsection{HS Vir\label{hs_sec}}

The Doppler fitting resulted in a dominant emission feature in the upper left
quadrant of the map (Fig.\ \ref{hsdop_fig}). However, both in strength and in 
shape there are significant residuals when comparing the reconstructed with 
the original data. Possible reasons are either an obscuration effect like
in VW Hyi (Section \ref{vwdop_sec}) or sub-orbital variations like
probably in AQ Eri (Section \ref{aqdop_sec}). A comparison of the line profiles
of the individual phase bins (Fig.\ \ref{hsdd_fig}) 
makes the second possibility appear
more probable. A good example is phase 0.15, where the
line profile over two orbits (left and right plot) has changed both in its 
general shape and even in the width of its wings. Furthermore, Howell et al.\
(\cite{howe+90}) found strong variations on non-orbital time scales in the
light curve of HS Vir, that could not be explained by flickering. In this case, 
Doppler mapping with respect to the orbital period yields no representative 
image of the emission distribution. It might be possible that this technique 
can be applied successfully, if a very large data set is used, so that 
non-orbital variations are averaged out.

\section{Discussion\label{disc_sec}}

\begin{table*}
\caption[]{Isolated emission features detected in the individual data sets,
which here appear sorted with increasing equivalent width. Cols.\ 2 to 4 give 
basic parameters of the systems, i.e.\ the period, the inclination,
the mass ratio $q = M_2/M_1$, and the (normal) outburst recurrence time, except
for RZ Leo and WX Cet, which show only superoutbursts. These parameters were 
taken from the Ritter \& Kolb (\cite{rittkolb98}) catalogue except where 
indicated. Cols.\ 5 and 6 contain the data set specific information on the 
equivalent width of the H$\alpha$ emission line and the accretion state during 
the observations (d = decline, q = quiescence), respectively. The 
$x$-velocities in Cols.\ 7 and 8 refer to the emission maxima in the 
respective halves of the Doppler maps, other emission regions are denoted by 
their identification. 
The presence of a feature is indicated by `$+$', its absence by
`$-$', uncertain ones by `?', and very uncertain ones by `??'.} 
\label{ies_tab}
\begin{tabular}{l l r r r r c c c c c l}
\hline\noalign{\smallskip}
data set & $P_{\rm orb}$ & $i$ & $q$ & $t_{\rm rec}$ & $W_{\rm H\alpha}$  
& state & $-v_x$ & $+v_x$ & secondary & gas & Fig. \\
 & [h] & [deg] & & [d] & [{\AA}] & & & & star & stream \\
\hline\noalign{\smallskip}
AK Cnc    & 1.62 &       36$^1$ &       0.28$^1$ &     47$^5$ & $-$28  
& d & $-$ & $-$ & $-$ & $-$ & \ref{akdop_fig} \\                       
HS Vir    & 1.85 & --           & 0.22$^6$       &          8 & $-$36  
& q & $+$ & $-$ & $-$ & $-$ & \ref{hsdop_fig}\\                        
VW Hyi 1  & 1.78 & $\sim$48$^7$ & $\sim$0.14$^7$ &         27 & $-$42  
& q & $-$ & $-$ & ?   & $+$ & \ref{vw1dop_fig}\\                       
VW Hyi 2  &      &              &                &            & $-$40  
& q & $+$ & $-$ & ??  & $+$ & \ref{vw2dop_fig}\\                       
VW Hyi 3  &      &              &                &            & $-$51  
& q & $+$ & $-$ & $-$ & $+$ & \ref{vw3dop_fig}\\                       
TU Men    & 2.81 & 52$^4$       & 0.46$^4$       &         37 & $-$118 
& q & $-$ & ??  & $+$ & $-$ & \ref{tudop_fig}\\                        
RZ Leo 1  & 1.84 & --           & --             & $>$365$^2$ & $-$139 
& q & $+$ & $+$ & $-$ & ??  & \ref{rz1dop_fig}\\                       
RZ Leo 2  &      &              &                &            & $-$118 
& q & $+$ & ?   & $-$ & ??  & \ref{rz2dop_fig}\\                       
AQ Eri 1a & 1.46 & --           & --             &         78 & $-$139 
& q & ?   & $-$ & ??  & ??  & \ref{aq1adop_fig}\\                      
AQ Eri 1b &      &              &                &            & $-$129 
& q & $+$ & $-$ & ??  & ??  & \ref{aq1bdop_fig}\\                      
AQ Eri 1c &      &              &                &            & $-$131 
& q & $+$ & $-$ & ??  & ??  & \ref{aq1cdop_fig}\\                      
WX Cet 1  & 1.40 &  $\le$65$^3$ &  $\ge$0.11$^3$ &        450 & $-$184 
& q & ?   & ?   & ?? & $-$  & \ref{wx1dop_fig} \\                      
WX Cet 2  &      &              &                &            & $-$198 
& q & ??  & $+$ & ?? & $-$  & \ref{wx2dop_fig}\\                       
WX Cet 3  &      &              &                &            & $-$183 
& q & $-$ & $+$ & ??  & $-$ & \ref{wx3dop_fig}\\
\hline\noalign{\smallskip}
\multicolumn{12}{l}{
References: 
1) Arenas \& Mennickent (\cite{arenmenn98}),
2) Ishioka et al.\ (\cite{ishi+01b}) ,
3) Mennickent (\cite{menn94}),}\\
\multicolumn{12}{l}{
4) Mennickent (\cite{menn95a}), 
5) Mennickent et al.\ (\cite{menn+96}),
6) Mennickent et al. (\cite{menn+99b}),
7) Tappert (\cite{tapp99})
}\\
\hline
\end{tabular}
\end{table*}

The systems analysed in this paper have in common that they are all of the SU 
UMa subclass, i.e.\ they are dwarf novae with orbital periods (in this case) 
ranging between 1.4 and 2.8 h, and they have low mass-transfer rates resulting
in comparatively faint, optically thin, accretion discs. Excluding the 
extremely long periodic system TU Men, the period range spanned by the systems 
even amounts
to only 27 min. It is therefore rather surprising that in these 
7 systems with supposedly common basic properties we are confronted with almost
the complete variety of isolated emission sources (IES).

One can now attempt to look for specific differences between the systems
regarding their parameters. In general, there appears to be a slight 
dependence for the IES on the inclination $i$ of the CV, in that emission from 
the trailing side of the disc 
($v_x < 0$; bright spot, etc.) is more likely to be present 
in high-inclination systems, and emission from the leading side is more 
frequent in low-inclination objects (Tappert \& Hanuschik \cite{tapphanu01}). 
Other possibly interesting parameters are the mass ratio $q = M_2/M_1$, which 
contains information on the stellar components, and the outburst recurrence 
time $t_\mathrm{rec}$, which is related to the transfer and accretion process.
Most of these parameters directly or indirectly affect the equivalent width 
$W_{\rm H\alpha}$ of the line. If the same mixture of effects is also 
responsible for the emission distribution, one should see a corresponding 
correlation. Finally, the orbital period $P_{\rm orb}$ is 
indicative of the mass of the secondary star.

We have listed these parameters together with the detected IES types in Table 
\ref{ies_tab}. With the range in $P_{\rm orb}$ being rather narrow, we do not 
expect it to play an important role in the IES distribution for the current 
sample. Instead the equivalent width seems to qualify as a better sorting 
criterion. It is influenced by a mixture of system (e.g.\ inclination) and 
disc (e.g.\ accretion rate) dependent parameters, which is likely to be 
also the case for the emission distribution. As all data sets for a specific 
system show in general similar additional emission components, we have sorted 
the data in Table \ref{ies_tab} according to the average equivalent width of a 
system. 

Still, there does not appear to exist any clear correlation with the 
IES type. For example, emission from the trailing 
side ($v_x < 0$) is observed 
almost over the whole range of $W_{\rm H\alpha}$, while emission from the 
secondary star (TU Men) is right in the middle of our sample. However, we note 
that emission from the leading side is only present in two systems with 
$W_{\rm H\alpha} > 100$ {\AA} (RZ Leo \& WX Cet). A look at the only other 
parameter that is roughly known for all the systems, the recurrence time, 
seemingly strengthens this tendency, since these two WZ Sge related systems 
represent the only ones that exclusively show superoutbursts. However this 
very probably is an artifical effect due to the smallness of our sample. IES on
the leading side have also been at least suspected\footnote{Doppler tomography
has been published for none of these systems, so that the evidence on their 
emission distribution is only secondary. This emphasises the need for more 
detailed line-profile studies for short period CVs.} e.g.\ in \object{SU UMa}
(Thorstensen et al.\ \cite{thor+86}) and \object{CY UMa} (Mart\'{\i}nez-Pais
\& Casares \cite{martcasa95}), which both show also normal outbursts and have
recurrence times much less than one year (Ritter \& Kolb \cite{rittkolb98}).

For the other parameters the list is largely incomplete. This is basically due 
to the fact that the calculation of $i$ and $q$ needs some knowledge about the 
secondary star, the latter being difficult to detect in short-period CVs. 
Furthermore, those parameters are usually rather uncertain with 1$\sigma$ 
errors in the order of 10 to 20\%.

An additional problem is that we are always dealing with a combination of
parameters, and a sample of 7 systems is certainly too small to provide
significant statistics in order to separate the individual influences.
So, while the variety of the 
phenomena detected here might come as a surprise,
the lack of a clear correlation with the system parameters probably should not.
   
Perhaps the only system whose emission distribution can be fixed to a parameter
is TU Men, which shows a strong contribution from the secondary star in its 
quiescent state. In SU UMa systems this phenomenon has been hitherto observed 
only in outburst. It appears that in this respect TU Men already qualifies as a
long-periodic system, whose disc is bright enough to induce emission on the 
surface of the secondary star.

Finally, we would like to point out that at least in one system the variations
between different data sets are not irregular. The analysis of spectra
from subsequent nights of VW Hyi show a clear long-term evolution from almost
pure gas-stream emission to a dominant bright spot. 
At the time of the observations, the system apparently is in a state of
a prolonged, slow decline (with a rate of roughly 1 mag in 40 days) about 20 
days after a normal outburst and $\sim$35 days before a superoutburst (Fig.\ 
\ref{vwlt_fig}). The long-term variations in the line-profile therfore take 
place during a period where the overall brightness of the system is barely 
changing.
Apparently even variations on small scales -- in the sense that they are not
accompanied or immediately followed by large photometric variations 
(i.e., outbursts) -- of the mass-transfer rate and/or in the disc itself yield 
significant changes in the line profile.

\section{Conclusions\label{concl_sec}}

The here presented atlas of line-profile analyses for 7 SU UMa-type dwarf novae
increases the number of available Doppler tomographies for such systems by
almost 100 \%. Our maps certainly suffer from the low quality of the data with 
respect to spectral resolution and low S/N. The reason for the former is that 
the data were not taken with the aim to perform this kind of analysis, 
but rather to 
derive the most basic parameter for CVs, the orbital period. Nevertheless, they
still served to gain a basic idea on the occurrence of isolated emission 
sources (IES) in these systems.
  
The noisy appearance of the maps, on the other hand, is mostly due to the fact 
that several systems showed non-orbital variations of the emission 
distribution, with the consequence that the respective data sets could not be 
combined in order to improve the S/N, thus actually representing one of the 
results of our study. Apart from that we can draw the following conclusions:

\begin{enumerate}
\item There is a large variety of IES phenomena present even in systems with
in principle very similar physical properties.
\item With HS Vir and probably also AQ Eri we find two systems which show
sub-orbital variations in their line profile. Consequently, Doppler tomography
does not yield a representative image of the emission distribution in these
cases. Very large data sets, where this type of variation can be averaged out,
might be needed for this purpose. These systems 
also represent interesting
targets for the recently introduced approach to map the emission line
flickering (Diaz \cite{diaz01}).
\item On the other hand, the non-orbital variations in VW Hyi occur on 
long-term timescales and therefore represent a trend in the evolution of the 
emission distribution. Here, long-term monitoring of the
inter-outburst state via Doppler mapping has the potential to provide important
clues on the outburst mechanisms (e.g.\ the role of irradiation).
\item TU Men can now claim the record of being the CV with the shortest
orbital period (and the only SU UMa dwarf nova) to show emission from the 
secondary star in quiescence.\footnote{Emission from the secondary star 
has possibly also been detected in the SU UMa dwarf nova \object{OU Vir} 
($P_\mathrm{orb}$ = 1.75 h) by Mason et al.\ (\cite{maso+02}). However, the 
authors are very cautious with their interpretation due to the problematic 
definition of the fiducial phase in this system.} Previously, this was the 
nova-like \object{DW UMa} ($P_{\rm orb}$ = 3.28 h), which showed this type of 
emission in its low state (Dhillon et al.\ \cite{dhil+94}).
\item The re-examination of `old' data with new techniques is important and 
meriting, even if only to motivate more detailed studies on the respective 
systems.
\end{enumerate}

\begin{acknowledgements}
We thank an anonymous referee for careful reading and thoughtful comments,
which helped to improve this paper.
We also thank Marcos Diaz for enlightening discussions on line-profile
variations. The VW Hyi data were taken by Walter Wargau ($\dagger$1996). 
RM acknowledges support by FONDECYT 1000324.
\end{acknowledgements}

\appendix

\newpage

\section{The atlas\label{atlas_sec}}
%%% AK Cnc %%%
\begin{figure}[h]
\resizebox{6.7cm}{!}{\includegraphics{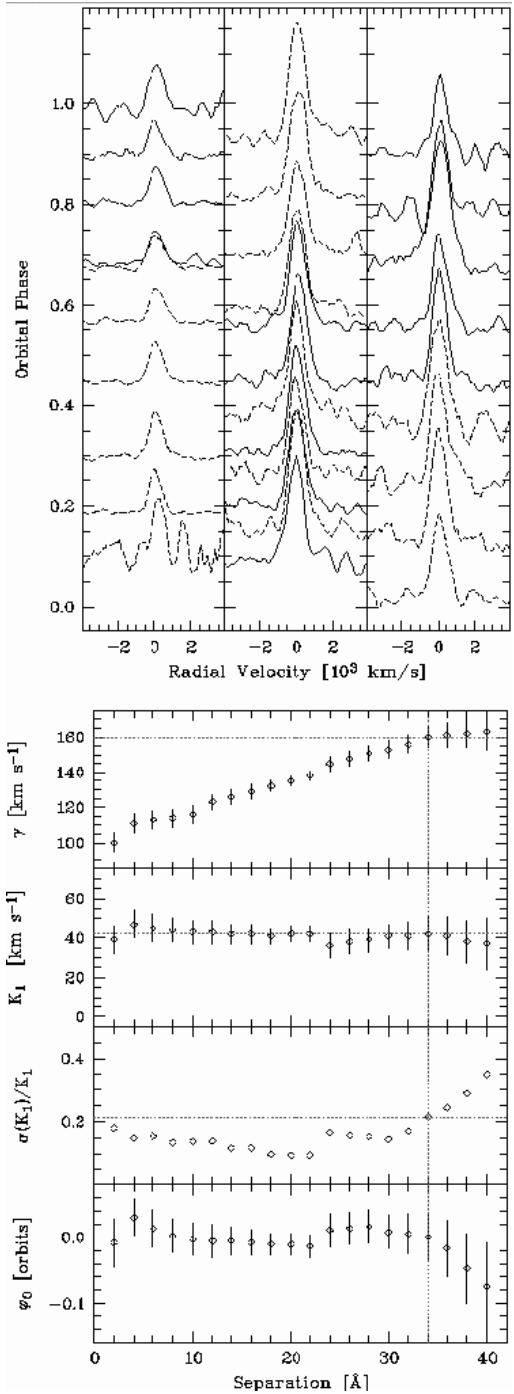}}
\caption{{\bf Top:} Line profiles of selected 10\% phase bins for AK Cnc. 
Data sets 1 -- 3 are plotted from left to right. The time sequence within one 
data set is symbolised by the sequence solid -- dashed. {\bf Bottom:}
Diagnostic diagram for AK Cnc, already corrected for the derived zero phase. 
The dotted lines mark the chosen separation and the corresponding
parameters.}
\label{akdd_fig}
\end{figure}
\begin{figure}
\begin{center}
\resizebox{7.5cm}{!}{\includegraphics{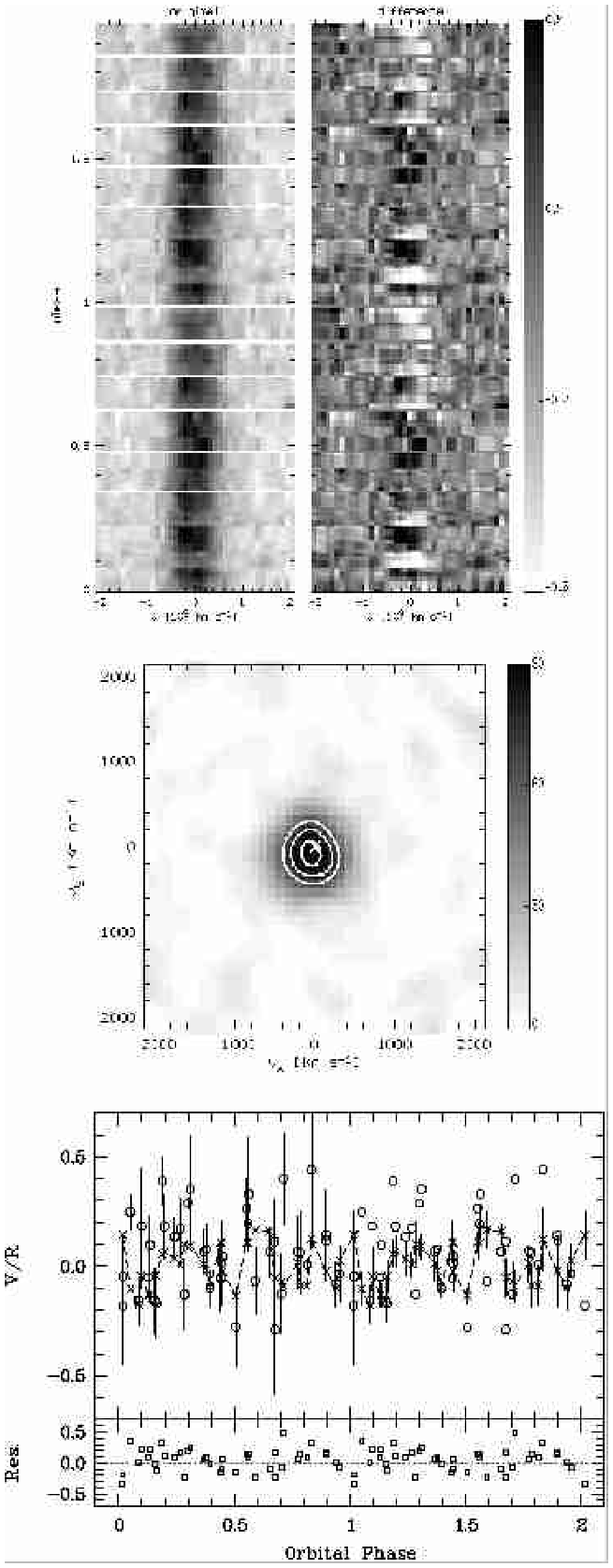}}
\end{center}
\caption{Doppler Tomography of AK Cnc. 
{\bf Top:} Original (left) and 
difference (original$-$reconstructed; right)
spectrum. The 
intensity bar on the right refers to the latter plot. 
{\bf Middle:} Doppler map. Contour levels are 
at 90, 75, and 60 per cent of the maximum intensity. 
{\bf Bottom:} $V/R$ plot of the original ($\circ$) and the reconstructed 
($\times$ and dashed line) data, and the residuals (bottom of the plot). In 
phases 0 to 1, error bars are given for the original data, phases 1 to 2 show 
those for the reconstructed data.}
\label{akdop_fig}
\end{figure}
%%% WX Cet %%%
\begin{figure}[h]
\resizebox{7.0cm}{!}{\includegraphics{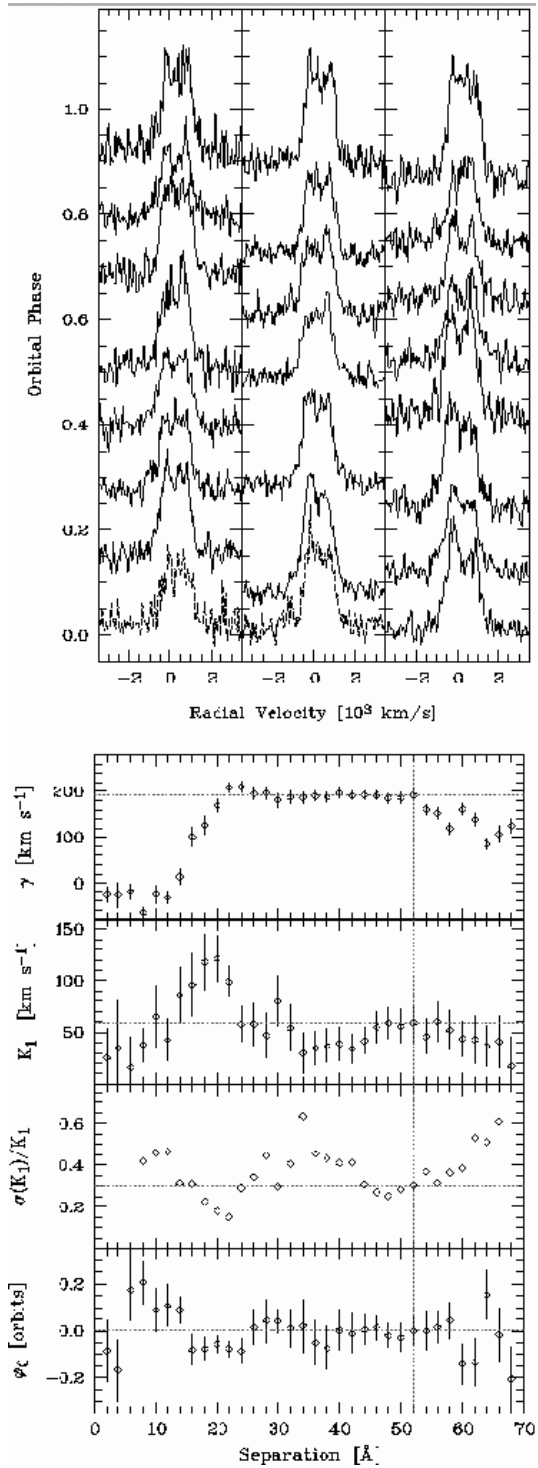}}
\caption{{\bf Top:} Line profiles of selected 10\% phase bins for WX Cet. 
Data sets 1 -- 3 are plotted from left to right. The time sequence within one 
data set is symbolised by the sequence solid -- dashed. {\bf Bottom:}
Diagnostic diagram for the combined data set of WX Cet, already corrected for 
the derived zero phase. 
The dotted lines mark the chosen separation and the corresponding
parameters.}
\label{wxdd_fig}
\end{figure}
\begin{figure}
\begin{center}
\resizebox{7.5cm}{!}{\includegraphics{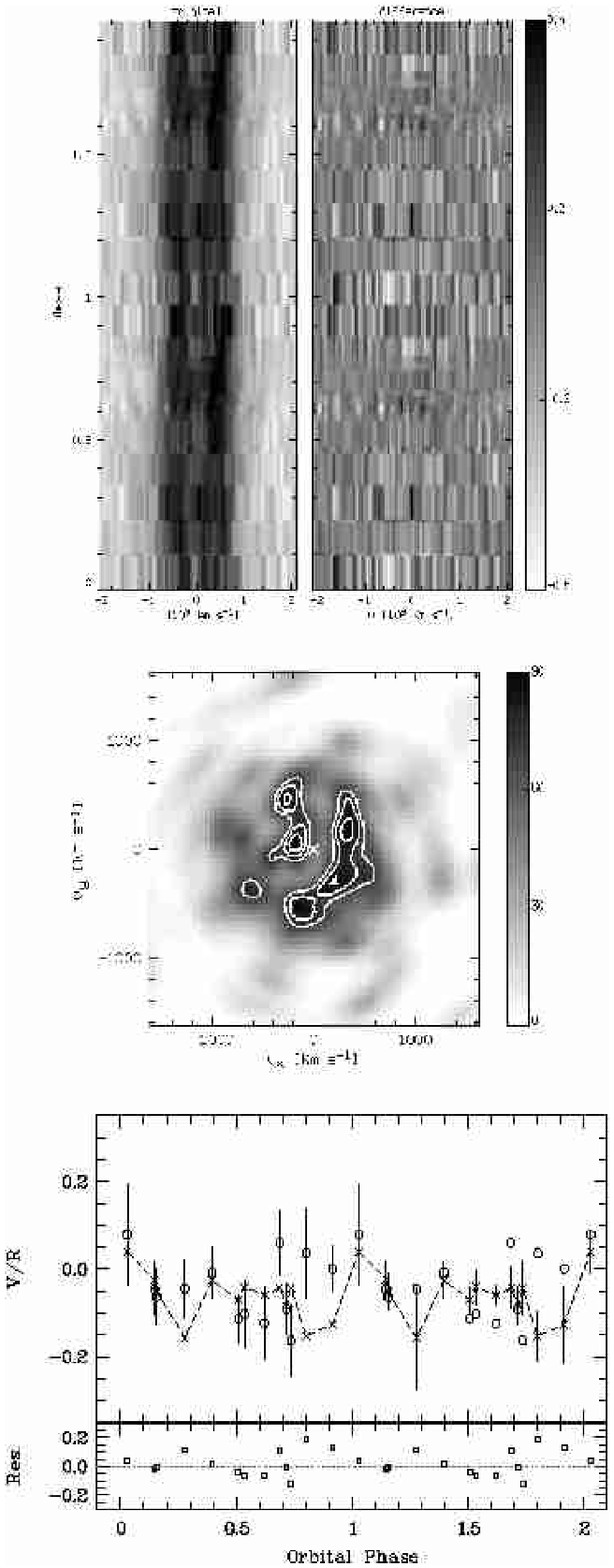}}
\end{center}
\caption{Doppler Tomography for WX Cet, data set 1. {\bf Top:} Original (left) 
and difference (original$-$reconstructed; right) spectrum. 
The intensity bar on the right refers to the latter plot. 
{\bf Middle:} Doppler map. Contour levels are 
at 86, 77, and 68 per cent of the maximum intensity. 
{\bf Bottom:} $V/R$ plot of the 
original ($\circ$) and the reconstructed ($\times$ and dashed line) data, and 
the residuals (bottom of the plot). In phases 0 to 1, error bars are given for 
the original data, phases 1 to 2 show those for the reconstructed data.}
\label{wx1dop_fig}
\end{figure}
\begin{figure}
\begin{center}
\resizebox{7.5cm}{!}{\includegraphics{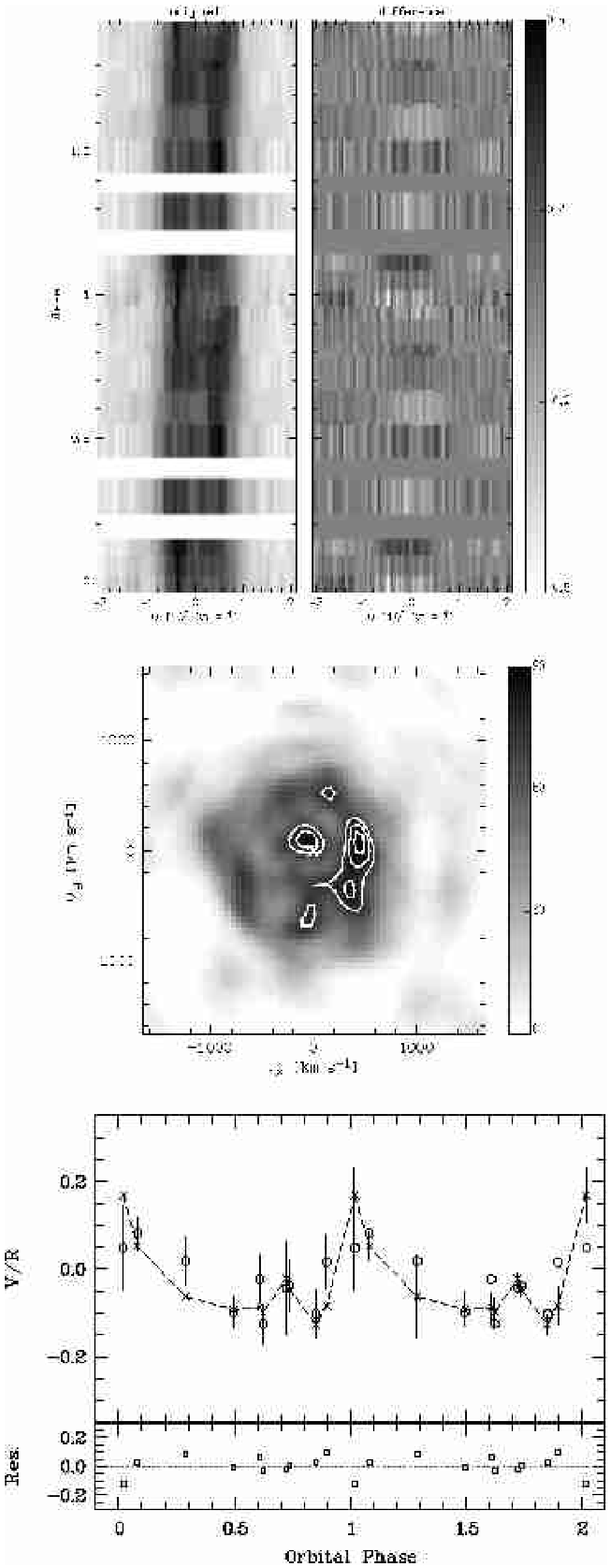}}
\end{center}
\caption{Doppler Tomography for WX Cet, data set 2. {\bf Top:} Original (left) 
and difference (original$-$reconstructed; right) spectrum. 
The intensity bar on the right refers to the latter plot. 
{\bf Middle:} Doppler map. Contour levels are 
at 90, 81, and 72 per cent of the maximum intensity. 
{\bf Bottom:} $V/R$ plot of the 
original ($\circ$) and the reconstructed ($\times$ and dashed line) data, and 
the residuals (bottom of the plot). In phases 0 to 1, error bars are given for 
the original data, phases 1 to 2 show those for the reconstructed data.}
\label{wx2dop_fig}
\end{figure}
\begin{figure}
\begin{center}
\resizebox{7.5cm}{!}{\includegraphics{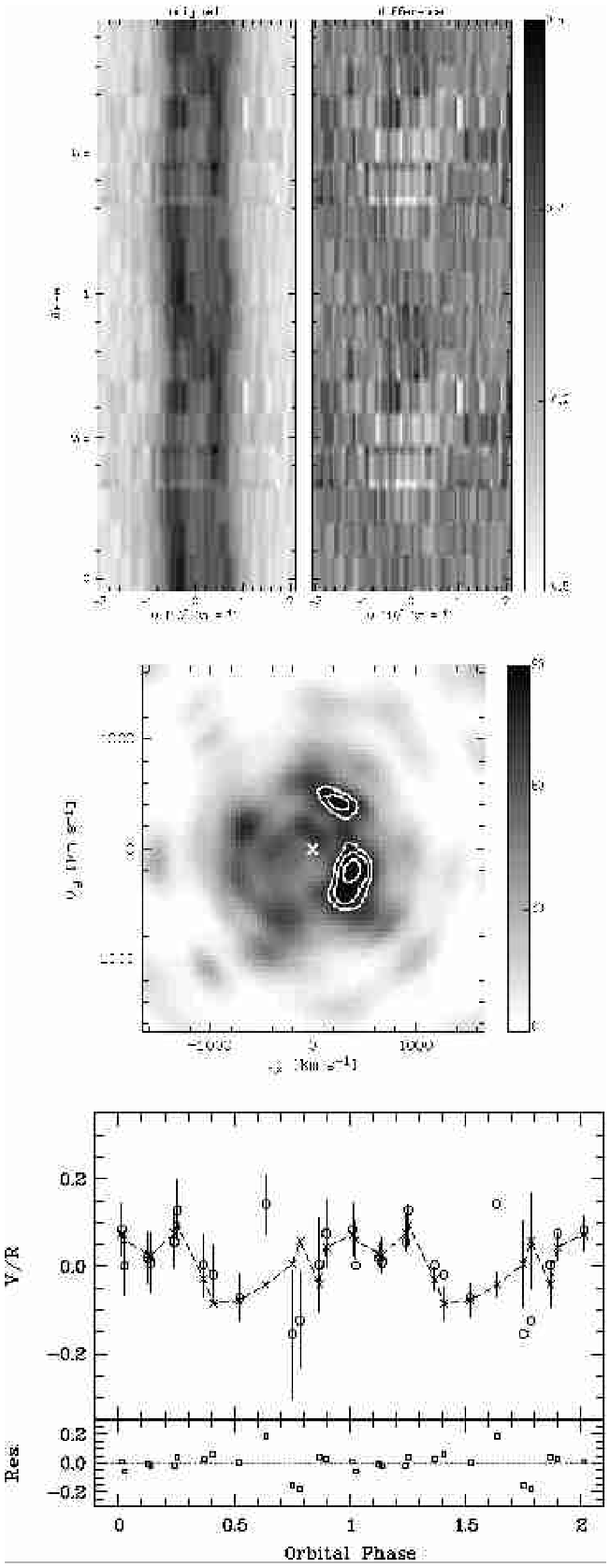}}
\end{center}
\caption{Doppler Tomography for WX Cet, data set 3. {\bf Top:} Original (left) 
and difference (original$-$reconstructed; right) spectrum. 
The intensity bar on the right refers to the latter plot. 
{\bf Middle:} Doppler map. Contour levels are 
at 93, 84, and 75 per cent of the maximum intensity. 
{\bf Bottom:} $V/R$ plot of the 
original ($\circ$) and the reconstructed ($\times$ and dashed line) data, and 
the residuals (bottom of the plot). In phases 0 to 1, error bars are given for 
the original data, phases 1 to 2 show those for the reconstructed data.}
\label{wx3dop_fig}
\end{figure}
%%% AQ Eri %%%
\begin{figure}[h]
\resizebox{7.0cm}{!}{\includegraphics{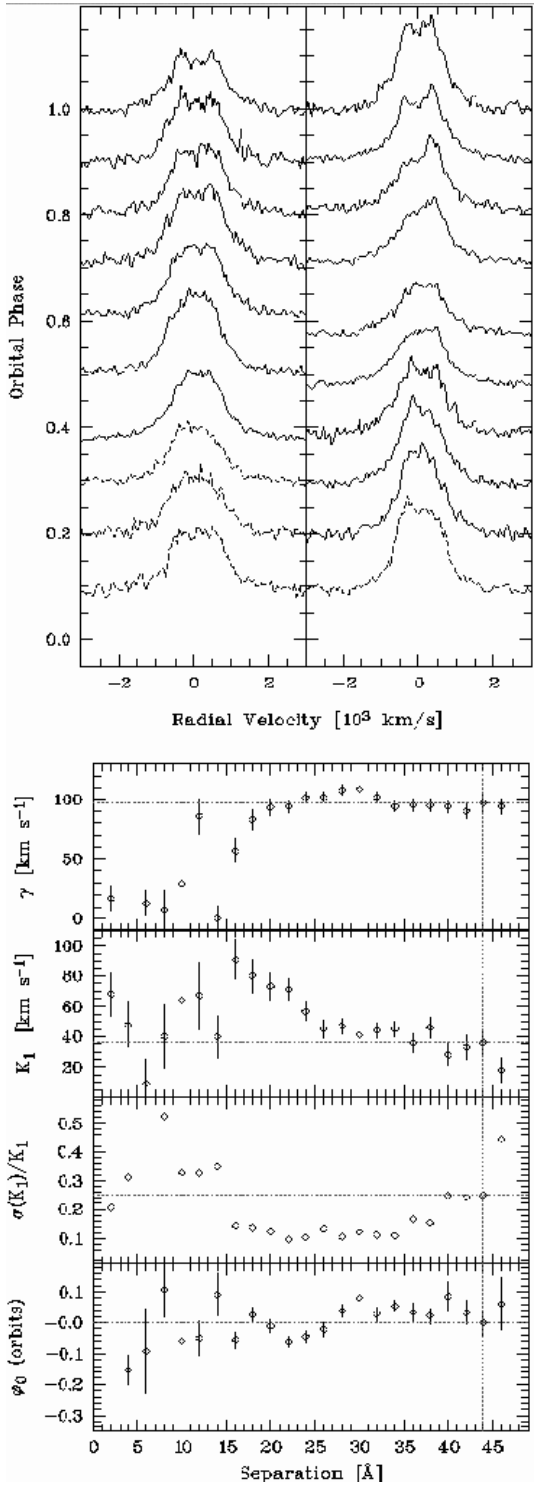}}
\caption{{\bf Top:} Line profiles of selected 10\% phase bins for AQ Eri. Data 
sets 1a and 1c are plotted on the left and on the right, respectively. The time
sequence within one data set is symbolised by the sequence solid -- dashed.
{\bf Bottom:} Diagnostic diagram for the combined data set of AQ Eri, already 
corrected for the derived zero phase. 
The dotted lines mark the chosen separation and the corresponding
parameters.}
\label{aqdd_fig}
\end{figure}
\begin{figure}
\begin{center}
\resizebox{7.5cm}{!}{\includegraphics{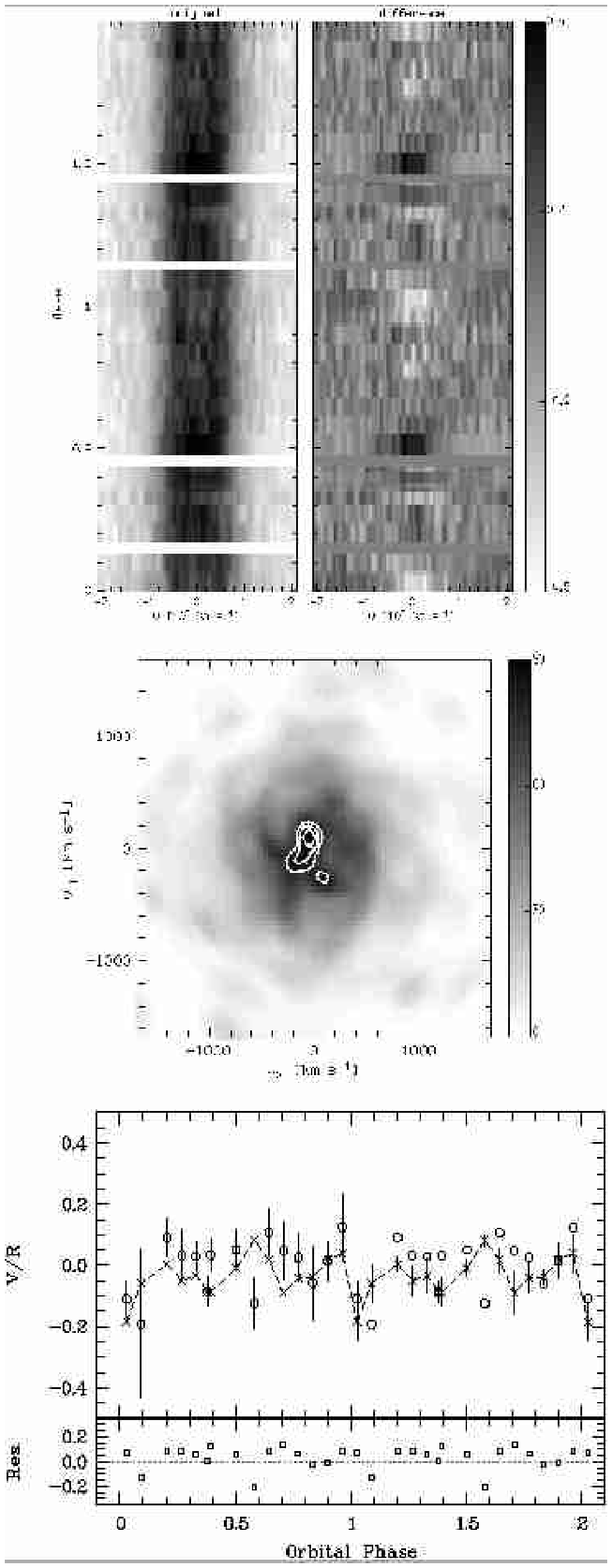}}
\end{center}
\caption{Doppler Tomography for AQ Eri, data set 1a. {\bf Top:} Original (left) 
and difference (original$-$reconstructed; right) spectrum. 
The intensity bar on the right refers to the latter plot. 
{\bf Middle:} Doppler map. Contour levels are 
at 90, 84, and 78 per cent of the maximum intensity. 
{\bf Bottom:} $V/R$ plot of the original ($\circ$) and the reconstructed 
($\times$ and dashed line) data, and the residuals (bottom of the plot). In 
phases 0 to 1, error bars are given for the original data, phases 1 to 2 show 
those for the reconstructed data.}
\label{aq1adop_fig}
\end{figure}
\begin{figure}
\begin{center}
\resizebox{7.5cm}{!}{\includegraphics{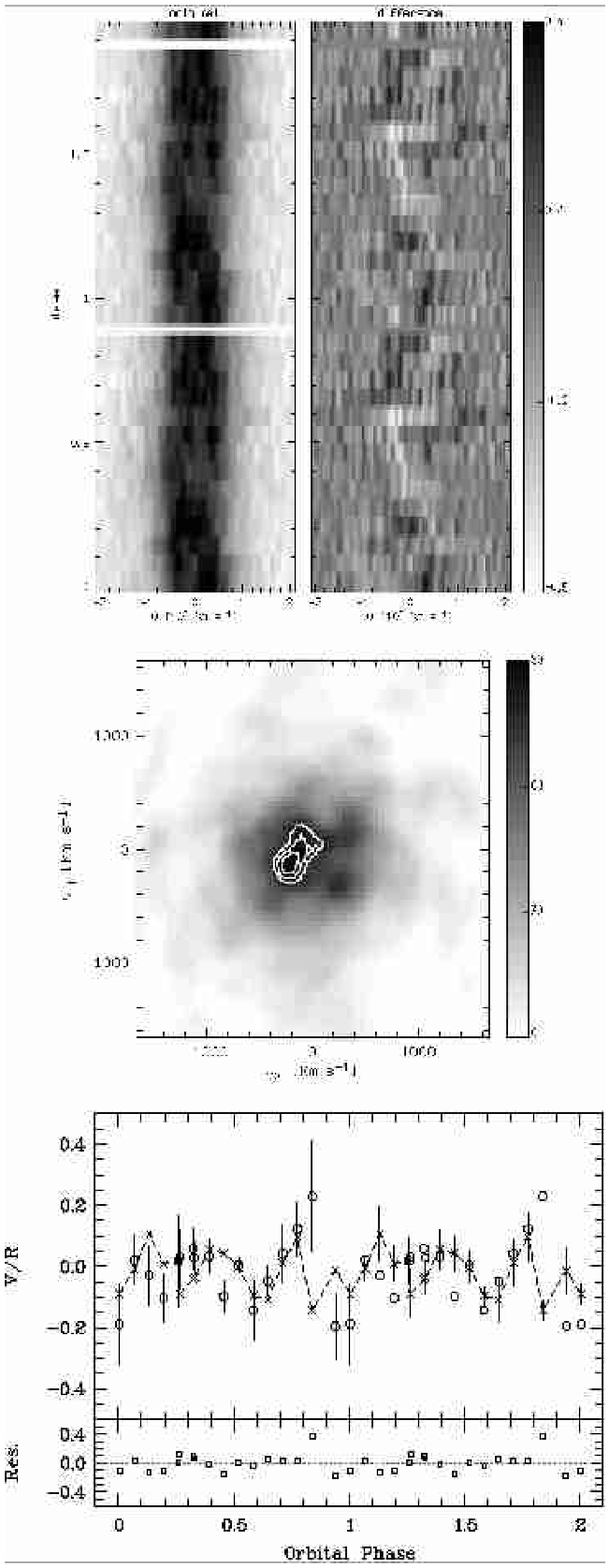}}
\end{center}
\caption{Doppler Tomography for AQ Eri, data set 1b. {\bf Top:} Original (left) 
and difference (original$-$reconstructed; right) spectrum. 
The intensity bar on the right refers to the latter plot. 
{\bf Middle:} Doppler map. Contour levels are 
at 90, 84, and 78 per cent of the maximum intensity. 
{\bf Bottom:} $V/R$ plot of the 
original ($\circ$) and the reconstructed ($\times$ and dashed line) data, and 
the residuals (bottom of the plot). In phases 0 to 1, error bars are given for 
the original data, phases 1 to 2 show those for the reconstructed data.}
\label{aq1bdop_fig}
\end{figure}
\begin{figure}
\begin{center}
\resizebox{7.5cm}{!}{\includegraphics{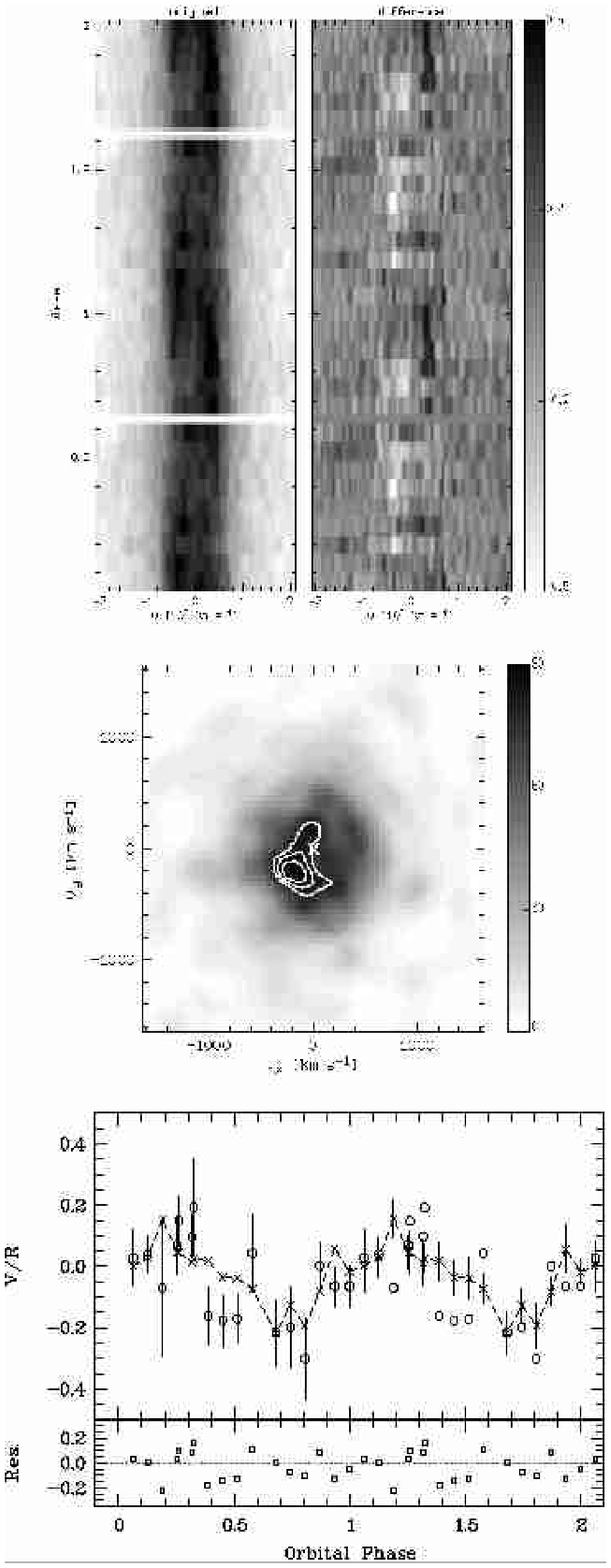}}
\end{center}
\caption{Doppler Tomography for AQ Eri, data set 1c. {\bf Top:} Original (left) 
and difference (original$-$reconstructed; right) spectrum. 
The intensity bar on the right refers to the latter plot. 
{\bf Middle:} Doppler map. Contour levels are 
at 90, 84, and 78 per cent of the maximum intensity. 
{\bf Bottom:} $V/R$ plot of the 
original ($\circ$) and the reconstructed ($\times$ and dashed line) data, and 
the residuals (bottom of the plot). In phases 0 to 1, error bars are given for 
the original data, phases 1 to 2 show those for the reconstructed data.}
\label{aq1cdop_fig}
\end{figure}
%%% VW Hyi %%%
\begin{figure}[h]
\resizebox{7.0cm}{!}{\includegraphics{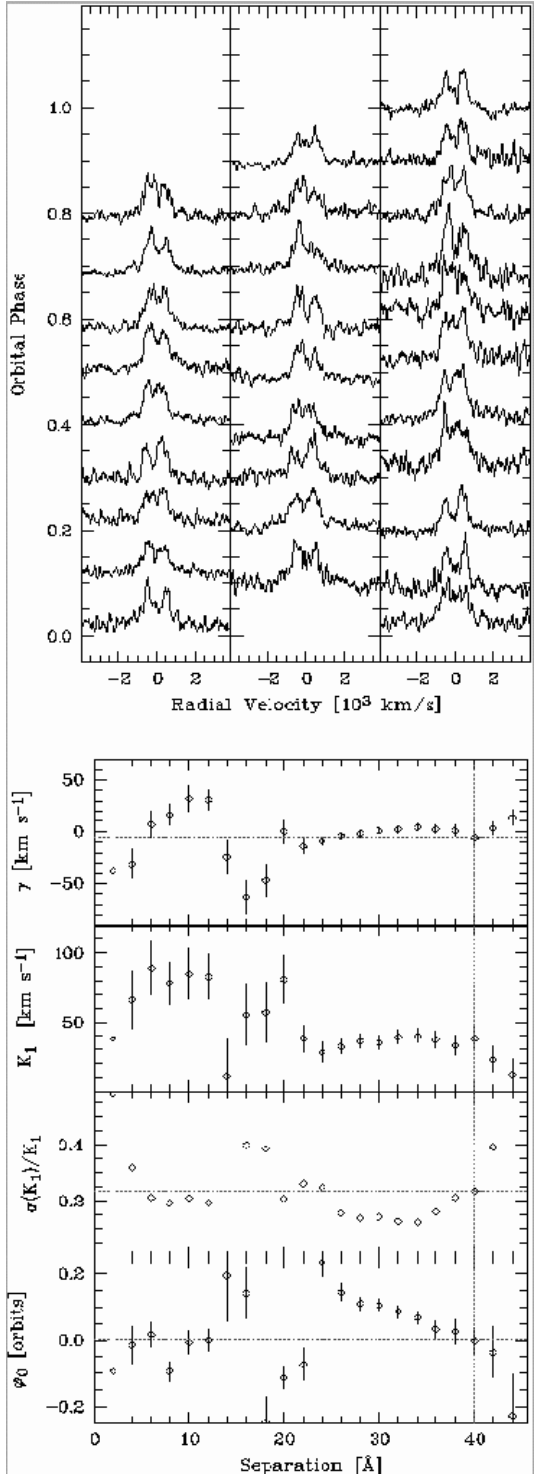}}
\caption{{\bf Top:} Line profiles of selected 10\% phase bins for VW Hyi.
Data sets 1 -- 3 are plotted from left to right. {\bf Bottom:} Diagnostic 
diagram for the combined data set of VW Hyi, already corrected for the derived 
zero phase. 
The dotted lines mark the chosen separation and the corresponding
parameters.}
\label{vwdd_fig}
\end{figure}
\begin{figure}
\begin{center}
\resizebox{7.5cm}{!}{\includegraphics{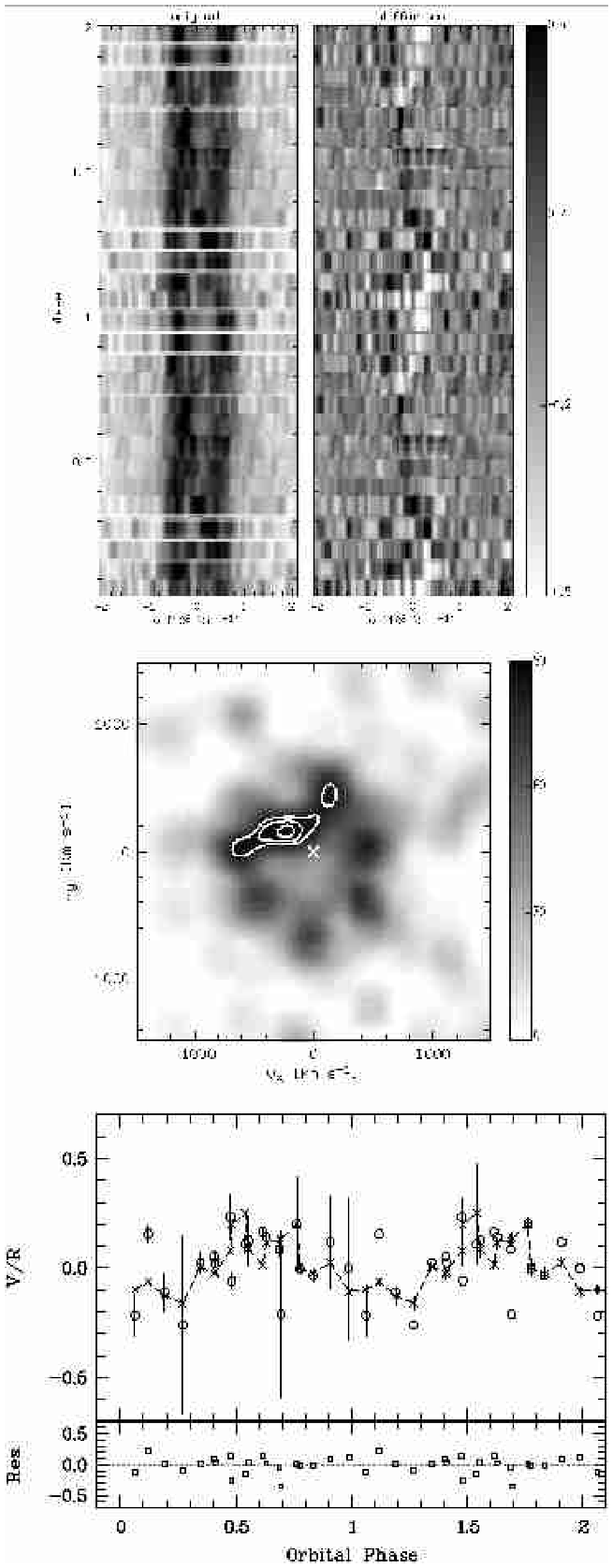}}
\end{center}
\caption{Doppler Tomography for VW Hyi, data set 1. {\bf Top:} Original (left) 
and difference (original$-$reconstructed; right)
spectrum.  The intensity bar on the right refers to the latter plot. 
{\bf Middle:} Doppler map. Contour levels are 
at 96, 89, and 82 per cent of the maximum intensity. 
{\bf Bottom:} $V/R$ plot of the 
original ($\circ$) and the reconstructed ($\times$ and dashed line) data, and 
the residuals (bottom of the plot). In phases 0 to 1, error bars are given for 
the original data, phases 1 to 2 show those for the reconstructed data.}
\label{vw1dop_fig}
\end{figure}
\begin{figure}
\begin{center}
\resizebox{7.5cm}{!}{\includegraphics{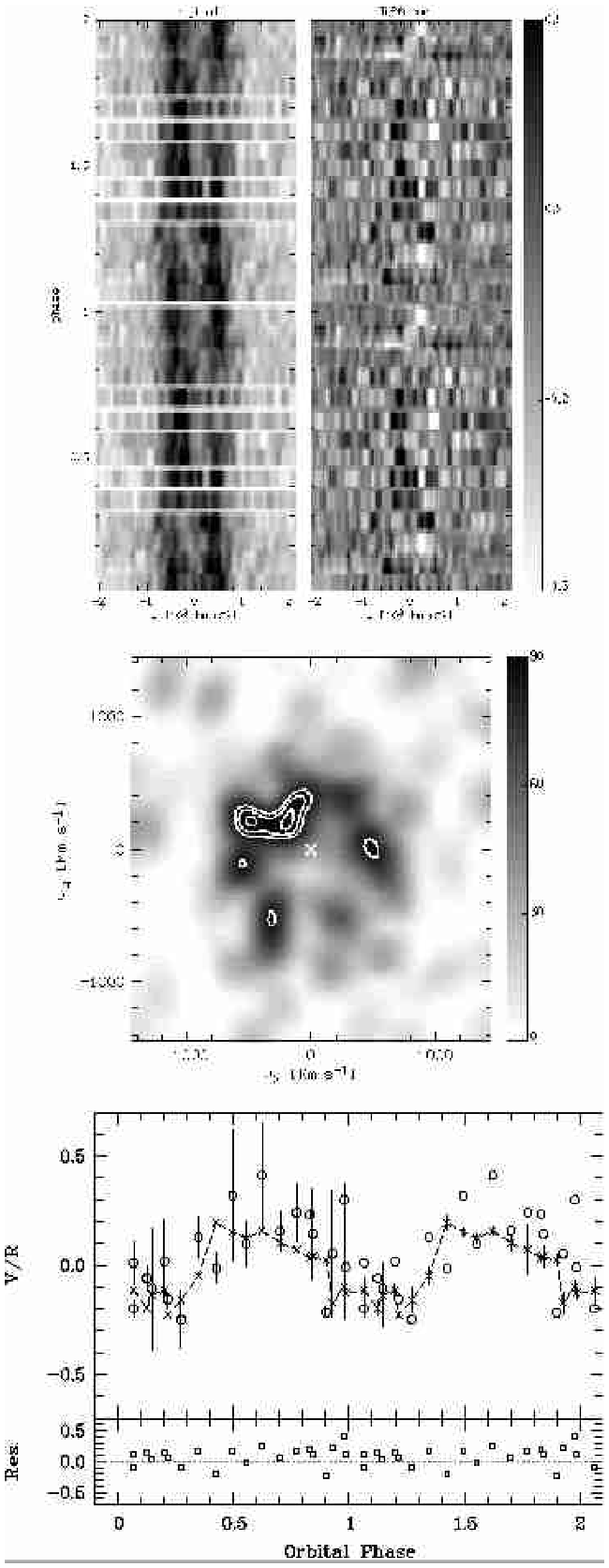}}
\end{center}
\caption{Doppler Tomography for VW Hyi, data set 2. {\bf Top:} Original (left) 
and difference (original$-$reconstructed; right) spectrum. 
The intensity bar on the right refers to the latter plot. 
{\bf Middle:} Doppler map. Contour levels are 
at 97, 90, and 83 per cent of the maximum intensity. 
{\bf Bottom:} $V/R$ plot of the 
original ($\circ$) and the reconstructed ($\times$ and dashed line) data, and 
the residuals (bottom of the plot). In phases 0 to 1, error bars are given for 
the original data, phases 1 to 2 show those for the reconstructed data.}
\label{vw2dop_fig}
\end{figure}
\begin{figure}
\begin{center}
\resizebox{7.5cm}{!}{\includegraphics{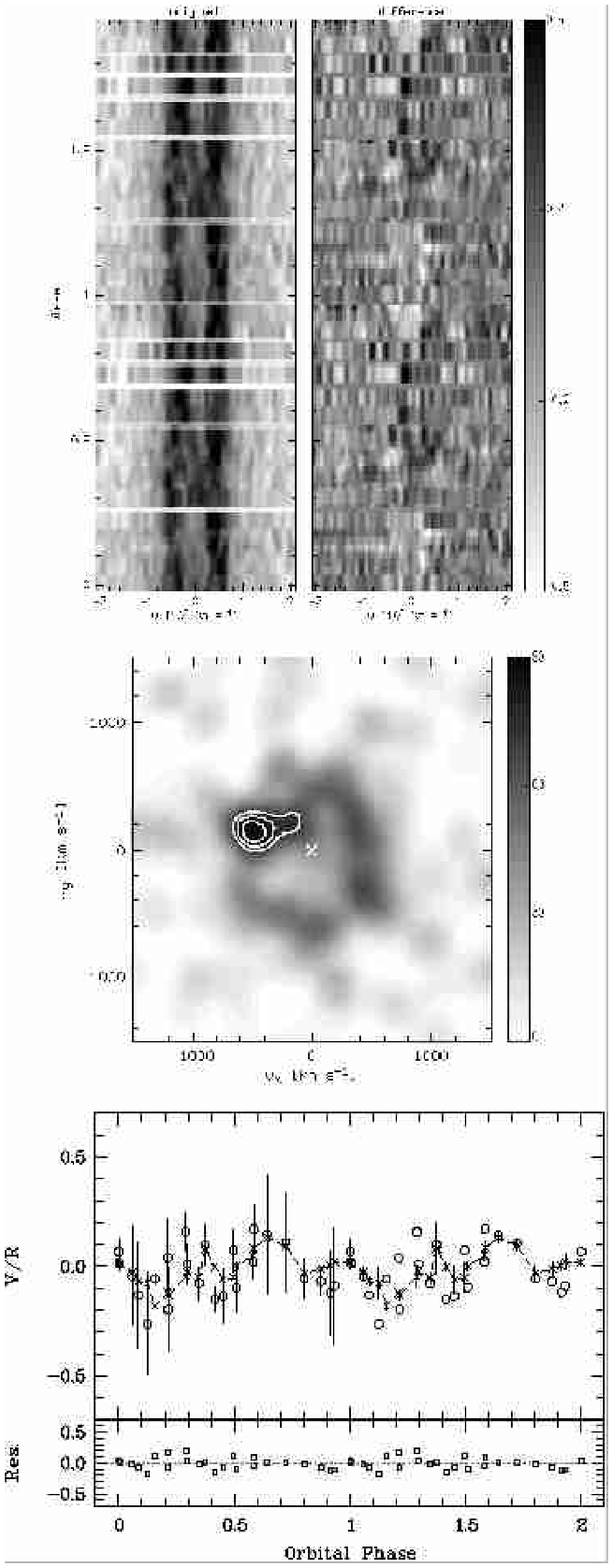}}
\end{center}
\caption{Doppler Tomography for VW Hyi, data set 3. {\bf Top:} Original (left) 
and difference (original$-$reconstructed; right) spectrum. 
The intensity bar on the right refers to the latter plot. 
{\bf Middle:} Doppler map. Contour levels are 
at 90, 80, and 70 per cent of the maximum intensity. 
{\bf Bottom:} $V/R$ plot of the 
original ($\circ$) and the reconstructed ($\times$ and dashed line) data, and 
the residuals (bottom of the plot). In phases 0 to 1, error bars are given for 
the original data, phases 1 to 2 show those for the reconstructed data.}
\label{vw3dop_fig}
\end{figure}
%%% RZ Leo %%%
\clearpage
\begin{figure}[h]
\resizebox{7.0cm}{!}{\includegraphics{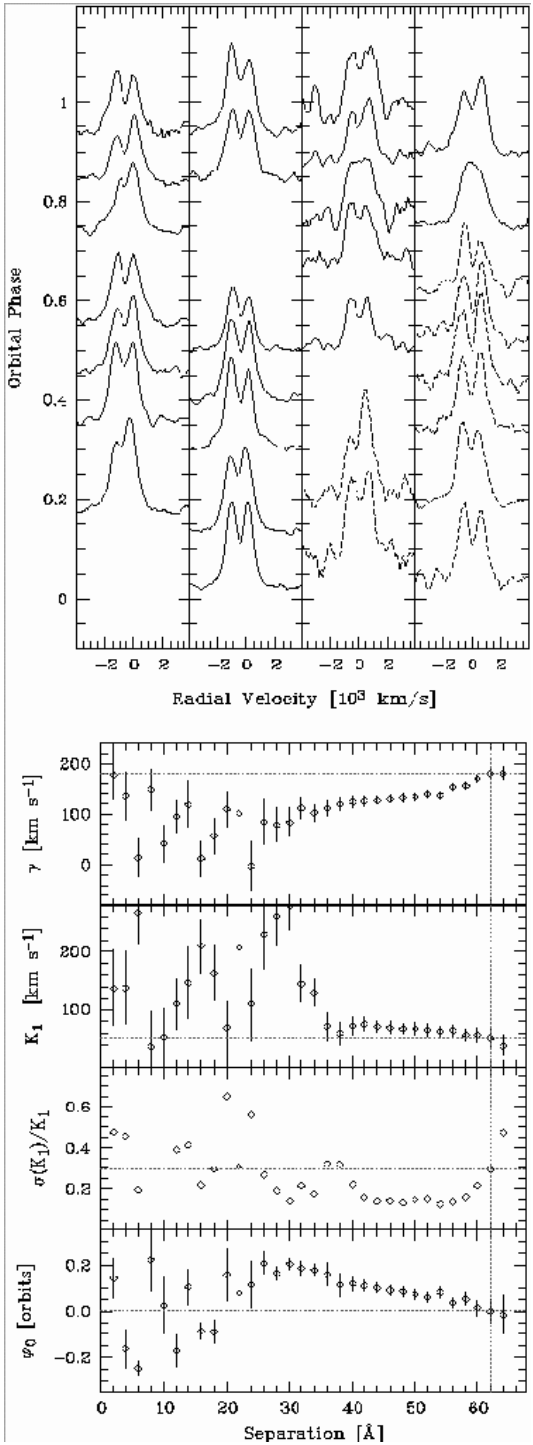}}
\caption{{\bf Top:} Line profiles of selected 10\% phase bins for RZ Leo.
Data sets 1a, 1b, 2a, 2b are plotted from left to right. The time sequence is 
symbolised by the line styles solid -- dashed. {\bf Bottom:} Diagnostic 
diagram of RZ Leo, data set 2, already corrected for the derived 
zero phase. 
The dotted lines mark the chosen separation and the corresponding parameters.}
\label{rzdd_fig}
\end{figure}
\begin{figure}
\begin{center}
\resizebox{7.5cm}{!}{\includegraphics{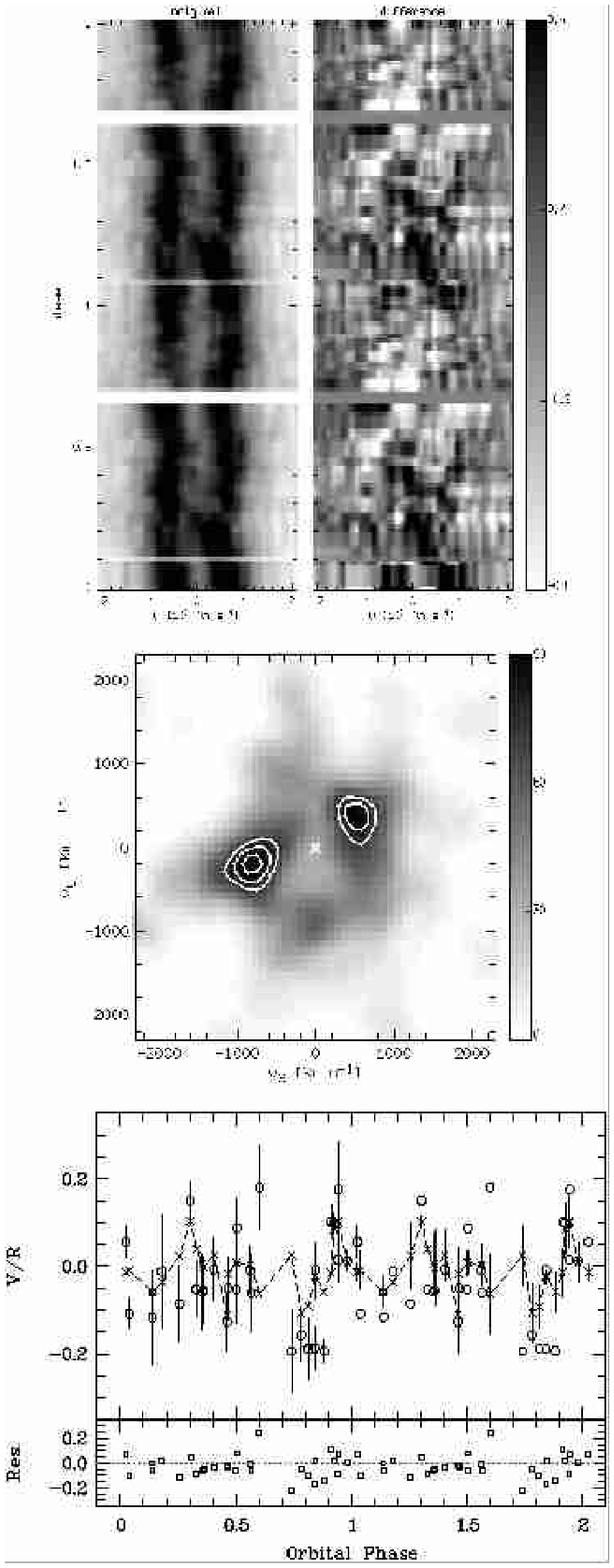}}
\end{center}
\caption{Doppler Tomography for RZ Leo, data set 1. {\bf Top:} Original (left) 
and difference (original$-$reconstructed; right) spectrum. 
The intensity bar on the right refers to the latter plot. 
{\bf Middle:} Doppler map. Contour levels are 
at 90, 80, and 70 per cent of the maximum intensity.
{\bf Bottom:} $V/R$ plot of the 
original ($\circ$) and the reconstructed ($\times$ and dashed line) data, and 
the residuals (bottom of the plot). In phases 0 to 1, error bars are given for 
the original data, phases 1 to 2 show those for the reconstructed data.}
\label{rz1dop_fig}
\end{figure}
\begin{figure}
\begin{center}
\resizebox{7.5cm}{!}{\includegraphics{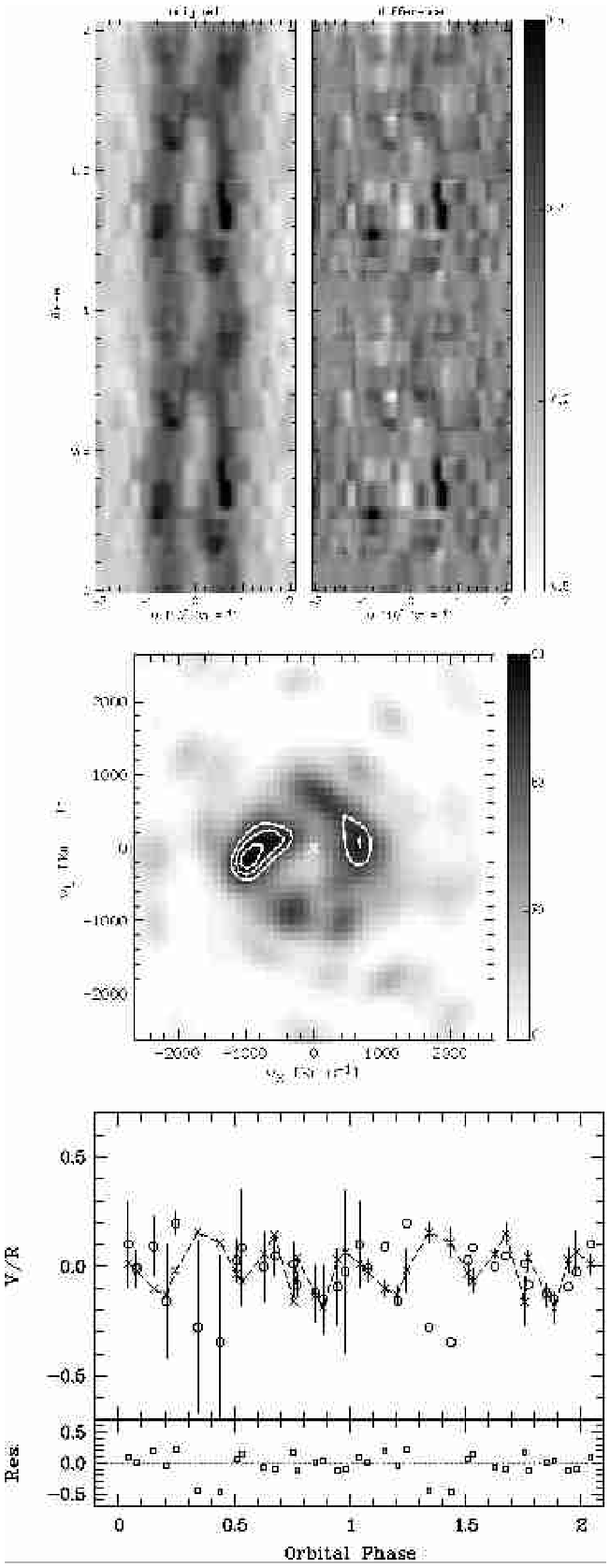}}
\end{center}
\caption{Doppler Tomography for RZ Leo, data set 2. {\bf Top:} Original (left) 
and difference (original$-$reconstructed; right) spectrum. 
The intensity bar on the right refers to the latter plot. 
{\bf Middle:} Doppler map. Contour levels are 
at 86, 76, and 66 per cent of the maximum intensity. 
{\bf Bottom:} $V/R$ plot of the 
original ($\circ$) and the reconstructed ($\times$ and dashed line) data, and 
the residuals (bottom of the plot). In phases 0 to 1, error bars are given for 
the original data, phases 1 to 2 show those for the reconstructed data.}
\label{rz2dop_fig}
\end{figure}
%%% TU Men %%%
\begin{figure}[h]
\resizebox{7.0cm}{!}{\includegraphics{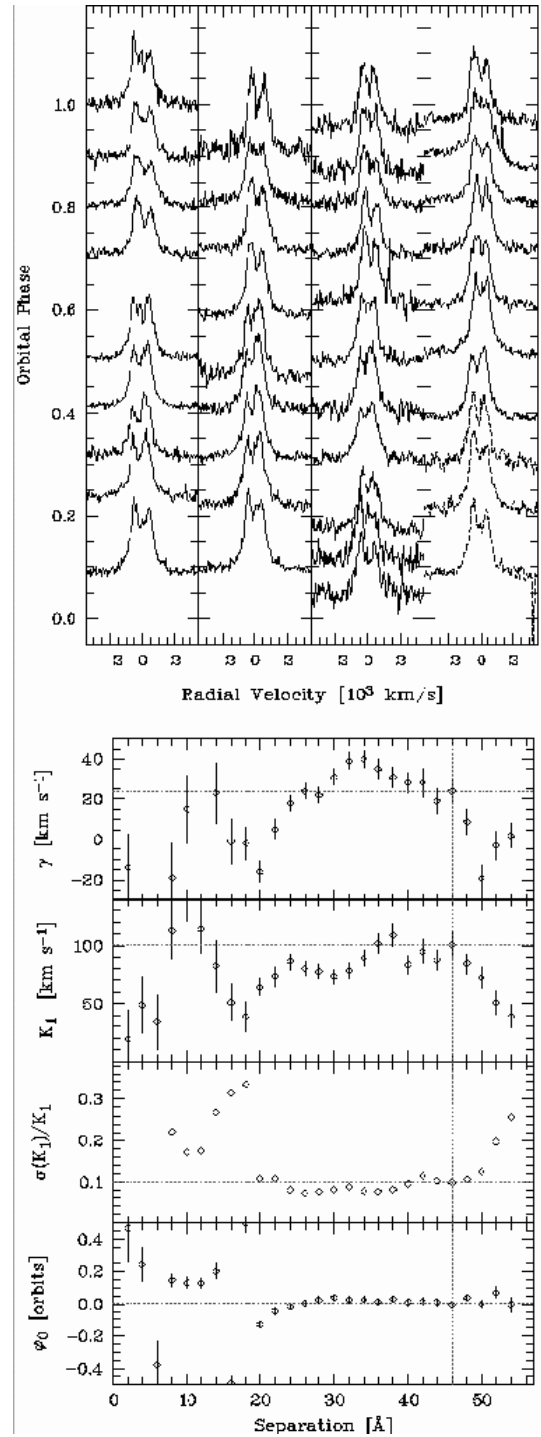}}
\caption{{\bf Top:} Line profiles of selected 10\% phase bins for TU Men.
Data sets 1 to 4 are plotted from left to right. The time sequence is 
symbolised by the line styles solid -- dashed. {\bf Bottom:} Diagnostic 
diagram for the combined data set of TU Men, already corrected for the derived 
zero phase.  
The dotted lines mark the chosen separation and the corresponding parameters.}
\label{tudd_fig}
\end{figure}
\begin{figure}
\begin{center}
\resizebox{7.5cm}{!}{\includegraphics{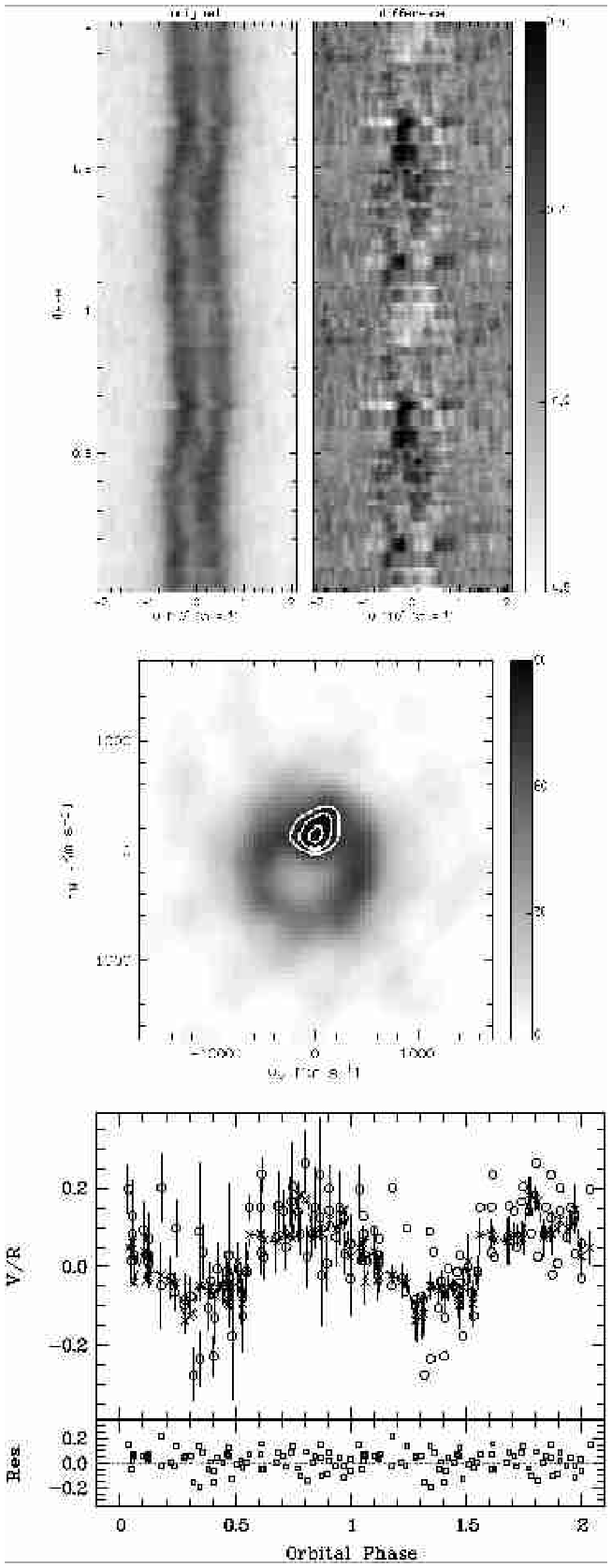}}
\end{center}
\caption{Doppler Tomography for TU Men. {\bf Top:} Original (left) 
and difference (original$-$reconstructed; right) spectrum. 
The intensity bar on the right refers to the latter plot. 
{\bf Middle:} Doppler map. Contour levels are 
at 94, 84, and 74 per cent of the maximum intensity. 
{\bf Bottom:} $V/R$ plot of the 
original ($\circ$) and the reconstructed ($\times$ and dashed line) data, and 
the residuals (bottom of the plot). In phases 0 to 1, error bars are given for 
the original data, phases 1 to 2 show those for the reconstructed data.}
\label{tudop_fig}
\end{figure}
%%% HS Vir %%%
%\clearpage
\begin{figure}[h]
\resizebox{7.0cm}{!}{\includegraphics{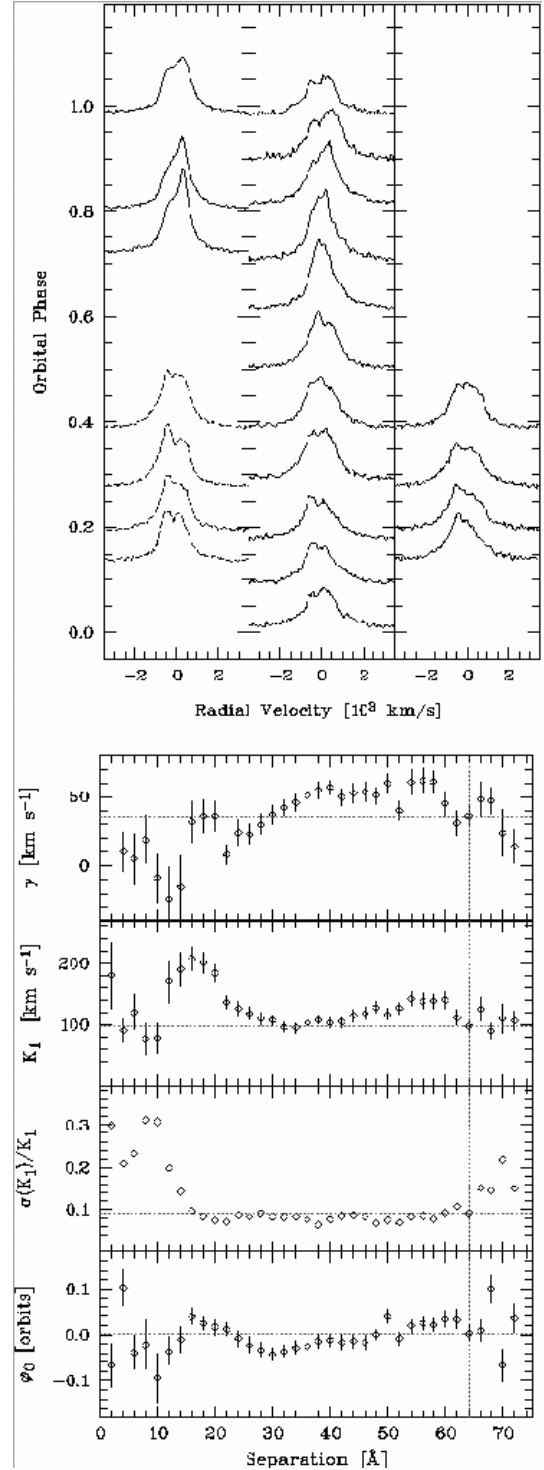}}
\caption{{\bf Top:} Line profiles of selected 10\% phase bins for HS Vir.
The time sequence in the left plot is symbolised by the sequence solid -- 
dashed. {\bf Bottom:} Diagnostic for HS Vir, already corrected for the derived 
zero phase.
The dotted lines mark the chosen separation and the corresponding parameters.}
\label{hsdd_fig}
\end{figure}
\begin{figure}
\begin{center}
\resizebox{7.5cm}{!}{\includegraphics{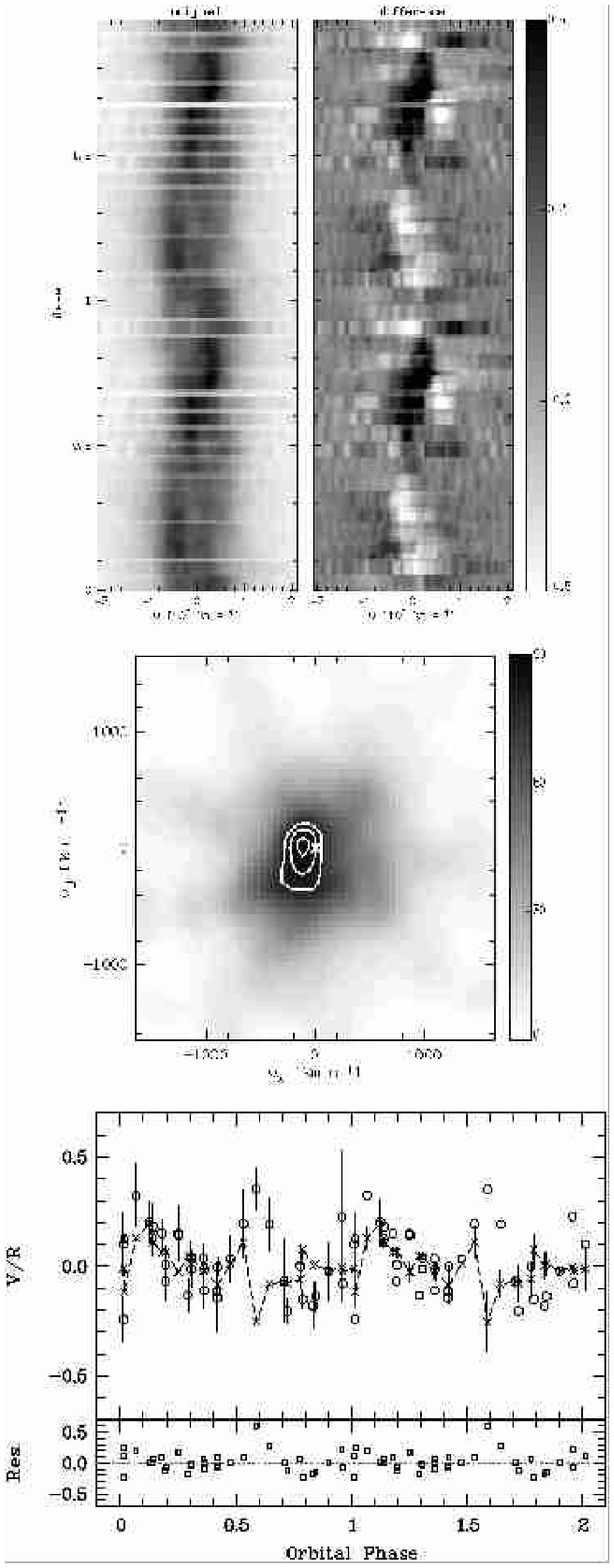}}
\end{center}
\caption{Doppler Tomography for HS Vir. {\bf Top:} Original (left) 
and difference (original$-$reconstructed; right) spectrum. 
The intensity bar on the right refers to the latter plot. 
{\bf Middle:} Doppler map. Contour levels are 
at 94, 86, and 78 per cent of the maximum intensity. 
{\bf Bottom:} $V/R$ plot of the 
original ($\circ$) and the reconstructed ($\times$ and dashed line) data, and 
the residuals (bottom of the plot). In phases 0 to 1, error bars are given for 
the original data, phases 1 to 2 show those for the reconstructed data.}
\label{hsdop_fig}
\end{figure}


\begin{thebibliography}{}

\bibitem[1998]{arenmenn98} 
Arenas, J. \& Mennickent, R.E. 1998, A\&A, 337, 472

\bibitem[1994]{dhil+94}
Dhillon, V.S., Jones, D.H.P. \& Marsh, T.R. 1994, MNRAS, 266, 859

\bibitem[2001]{diaz01}
Diaz, M.P., 2001, ApJL, 553, L177

\bibitem[2000]{hoar+00}
Hoard, D.W., Thorstensen, J.R. \& Szkody, P. 2000, ApJ, 537, 936

\bibitem[1995]{horn95}
Horne, K., 1995, A\&A, 297, 273

\bibitem[1986]{hornmars86}
Horne, K. \& Marsh, T.M. 1986, MNRAS, 218, 761

\bibitem[1990]{howe+90}
Howell, S.B., Szkody, P., Kreidel, T.J., Mason, K.O. \& Puchnarewicz, E.M. 
1990, PASP, 102, 758

\bibitem[2001]{ishi+01b}
Ishioka, R., Kato, T., Uemura, M., et al. 2001, PASJ, 53, 905

\bibitem[1994]{kait+94}
Kaitchuck, R.H., Schlegel, E.M., Honeycutt, R.K., et al. 1994, ApJS, 93, 519

\bibitem[1998]{kait+98}
Kaitchuck, R. H., Schlegel, E. M., White II, J. C. \& Mansperger, C. S. 1998, 
ApJ, 499, 444

\bibitem[2001]{mars01}
Marsh, T.R. 2001, in: {\em Astrotomography, Indirect Imaging Methods in
Observational Astronomy}, ed. H.M.J. Boffin, D. Steeghs \& J. Cuypers, Lect.
Notes on Phys., 573, 1

\bibitem[1988]{marshorn88}
Marsh, T.R. \& Horne, K. 1988, MNRAS, 235, 269

\bibitem[1995]{martcasa95}
Mart\'{\i}nez-Pais, I.G. \& Casares, J. 1995, MNRAS, 275, 699

\bibitem[1994]{mart+94}
Mart\'{\i}nez-Pais, I.G., Giovannelli, F., Rossi, C. \& Gaudenzi, S. 1994, 
A\&A, 291, 455

\bibitem[1996]{mart+96}
Mart\'{\i}nez-Pais, I.G., Giovannelli, F., Rossi, C. \& Gaudenzi, S. 1996, 
A\&A, 308, 833

\bibitem[2002]{maso+02}
Mason, E., Howell, S.B., Szkody, P., et al. 2002, A\&A, 396, 633

\bibitem[1999]{matt99}
Mattei, J.A. 1999, Observations from the AAVSO International Database, private
communication

\bibitem[1994]{menn94}
Mennickent, R.E. 1994, A\&A, 285, 979

\bibitem[1995a]{menn95a}
Mennickent, R.E. 1995a, A\&A, 294, 126

\bibitem[2001]{menntapp01}
Mennickent, R.E. \& Tappert, C. 2001, A\&A, 372, 563

\bibitem[1995b]{menn95b}
Mennickent, R.E. 1995b, Ph.D. Thesis, Pontificia Universidad Cat\'olica de 
Chile

\bibitem[1996]{menn+96}
Mennickent, R.E., Nogami, D., Kato, T. \& Worraker, W. 1996, A\&A, 315, 493

\bibitem[1999]{menn+99b}
Mennickent, R.E., Matsumoto, K. \& Arenas, J. 1999, A\&A, 348, 466

%\bibitem[2002]{menn+02}
%Mennickent, R.E., Tappert, C. \& Diaz, M., 2002, ESO Mess., 109, 41

\bibitem[2001]{nort+01}
North, R., Marsh, T.R., Moran, C.K.J., et al. 2001, in: {\em Astrotomography, 
Indirect Imaging Methods in Observational Astronomy}, ed. H.M.J. Boffin, D. 
Steeghs, \& J. Cuypers, Lect. Notes on Phys., Vol. 573, p.33

\bibitem[1998]{rittkolb98}
Ritter, H. \& Kolb, U. 1998, A\&AS, 129, 83

\bibitem[1980]{schnyoun80}
Schneider, D.P. \& Young, P. 1980, ApJ, 238, 946

\bibitem[1983a]{shaf83a}
Shafter, A.W. 1983a, ApJ, 267, 222

\bibitem[1983b]{shaf83b}
Shafter, A.W. 1983b, Ph.D. Thesis, University of California, Los Angeles

\bibitem[2000]{skid+00}
Skidmore, W., Mason, E., Howell, S.B., et al. 2000, MNRAS, 318, 429

\bibitem[1981]{smak81b}
Smak, J. 1981, Acta Astron., 31, 395

\bibitem[1998]{spru98}
Spruit, H.C. 1998, astro-ph/9806141

\bibitem[2003]{stee03}
Steeghs, D. 2003, MNRAS, in press (astro-ph/0305365)

\bibitem[1997]{stee+97}
Steeghs, D., Harlaftis, E.T. \& Horne, K. 1997, MNRAS, 290, L28

\bibitem[1981]{stov81a}
Stover, R.J. 1981, ApJ, 248, 684

\bibitem[1999]{tapp99}
Tappert, C. 1999, Ph.D. Thesis, Ruhr-Universit\"at Bochum

\bibitem[2001]{tapphanu01}
Tappert, C. \& Hanuschik, R.W. 2001, in: {\em Astrotomography, Indirect 
Imaging Methods in Observational Astronomy}, ed. H.M.J. Boffin, D. Steeghs \& 
J. Cuypers, Lect. Notes on Phys., 573, 119

\bibitem[1986]{thor+86}
Thorstensen, J.R., Wade, R.A. \& Oke, J.B. 1986, ApJ, 309, 721

\bibitem[1995]{warn95}
Warner, B. 1995, {\sl Cataclysmic Variable Stars}, Cambridge University Press


\end{thebibliography}
\end{document}